\documentclass{tlp}
\usepackage{amsmath}
\usepackage{graphicx}

  \title[Expressiveness of Communication in ASP]{Expressiveness of Communication \\in Answer Set Programming}

  \author[K. Bauters et al.]
         {KIM BAUTERS, STEVEN SCHOCKAERT\\
         Dept. of Applied Mathematics and Computer Science,\\ Krijgslaan 281 (WE02), Universiteit Gent, 9000 Gent, Belgium\\
         \email{(kim.bauters, steven.schockaert)@ugent.be}
         \and
         JEROEN JANSSEN, DIRK VERMEIR\\
         Dept. of Computer Science, Vrije Universiteit Brussel,\\ Pleinlaan 2, 1050 Brussel, Belgium\\
         \email{(jeroen.janssen, dirk.vermeir)@vub.ac.be}
         \and
         MARTINE DE COCK\\
         Dept. of Applied Mathematics and Computer Science,\\ Krijgslaan 281 (WE02), Universiteit Gent, 9000 Gent, Belgium\\
         \email{martine.decock@ugent.be}
         }
         
\makeatletter
\newenvironment{oneshot}[1]{\@begintheorem{#1}{\unskip}}{\@endtheorem} 
\makeatother

\DeclareSymbolFont{lettersA}{U}{txmia}{m}{it}
\DeclareMathSymbol{\natural}{\mathord}{lettersA}{"8E}

\newtheorem{definition}{Definition}
\newtheorem{example}{Example}
\newtheorem{lemma}{Lemma}
\newtheorem{proposition}{Proposition}
\newtheorem{corollary}{Corollary}

\newcommand{\fresh}[1]{{#1}^{\dagger}}

\newcommand{\eg}[0] {\emph{e.g.}~}
\newcommand{\ie}[0] {\emph{i.e.}~}
\newcommand{\wrt}[0] {\emph{w.r.t.}~}
\newcommand{\Ref}[1] {(\ref{#1})}
\def\us{\char`\_}

\newcommand{\themydef}[1]{Definition~\ref{def:#1}}
\newcommand{\themyex}[1]{Example~\ref{ex:#1}}
\newcommand{\themyprop}[1]{Proposition~\ref{prop:#1}}
\newcommand{\themylem}[1]{Lemma~\ref{lem:#1}}
\newcommand{\themycor}[1]{Corollary~\ref{cor:#1}}
\newcommand{\thesec}[1]{Section~\ref{sec:#1}}
\newcommand{\thefig}[1]{Figure~\ref{fig:#1}}

\newcommand{\naf}[0]{not~}
\newcommand{\snaf}[0]{not}

\newcommand{\arule}[2]{\ensuremath{#1 \leftarrow #2}}
\newcommand{\sarule}[2]{\ensuremath{\mathit{#1} &\leftarrow \mathit{#2}}}

\newcommand{\hbase}[1]{\ensuremath{\mathcal{B}_{#1}}}
\newcommand{\struprog}[1]{\ensuremath{\mathcal{#1}}}
\newcommand{\proj}[2]{{#1}_{ #2}}
\newcommand{\ask}[2]{\ensuremath{{#1\!:\!#2}}}
\newcommand{\sask}[2]{\ensuremath{{#1\us#2}}}

\newcommand{\aset}[1] {\ensuremath{\left\{#1\right\}}}

\newcommand{\vect}[1] {\ensuremath{\left(#1\right)}}
\providecommand{\set}[1] {\aset{#1}}
\newcommand{\card}[1]{{\ensuremath{\mid\!{#1}\!\mid}}}

\newcommand{\complexity}[1]{\ensuremath{\mathsf{#1}}}
\newcommand{\cC}[0] {\complexity{C}}
\newcommand{\cP}[0] {\complexity{P}}
\newcommand{\cNP}[0]{\complexity{NP}}
\newcommand{\cco}[0]{\complexity{co}}
\newcommand{\ccoNP}[0]{\complexity{coNP}}
\newcommand{\hisig}[2]{\ensuremath{\Sigma_{#1}^{#2}}}
\newcommand{\hipi}[2]{\ensuremath{\Pi_{#1}^{#2}}}
\newcommand{\hidelta}[2]{\ensuremath{\Delta_{#1}^{#2}}}
\newcommand{\spolpi}[1]{\hipi{#1}{\mathbf{\cP}}}
\newcommand{\spoldelta}[1]{\hidelta{#1}{\mathbf{\cP}}}
\newcommand{\spolsig}[1]{\hisig{#1}{\mathbf{\cP}}}

\newcommand{\condset}[2]{\aset{{#1}\mid{#2}}}

\newcommand{\range}[2] {\rangec{#1}{#2}{,}}
\newcommand{\rangec}[3] {\ensuremath{#1#3\ldots#3#2}}
\newcommand{\srange}[3] {\range{#1_{#2}}{#1_{#3}}}
\newcommand{\srangec}[4] {\rangec{#1_{#2}}{#1_{#3}}{#4}}

\providecommand{\suchthat}{\ensuremath{ \cdot }}
\newcommand{\sForall}[2]{\ensuremath{\forall #1 \suchthat #2}}

\newcommand{\concept}[1]{\emph{#1}}

\newenvironment{reproposition}[1]{
\begin{oneshot}{\themyprop{#1}}
}{
\end{oneshot}
}

\newenvironment{relemma}[1]{
\begin{oneshot}{\themylem{#1}}
}{
\end{oneshot}
}

\submitted{23 February 2011}
\revised{26 August 2011}
\accepted{5 September 2011}

\begin{document}

\maketitle

\label{firstpage}

  \begin{abstract}
  Answer set programming (ASP) is a form of declarative programming that allows to succinctly formulate and efficiently solve complex problems. An intuitive extension of this formalism is communicating ASP, in which multiple ASP programs collaborate to solve the problem at hand. However, the expressiveness of communicating ASP has not been thoroughly studied. In this paper, we present a systematic study of the additional expressiveness offered by allowing ASP programs to communicate. First, we consider a simple form of communication where programs are only allowed to ask questions to each other. For the most part, we deliberately only consider simple programs, \ie programs for which computing the answer sets is in~$\cP$. We find that the problem of deciding whether a literal is in some answer set of a communicating ASP program using simple communication is \cNP-hard. In other words: we move up a step in the polynomial hierarchy due to the ability of these simple ASP programs to communicate and collaborate. Second, we modify the communication mechanism to also allow us to focus on a sequence of communicating programs, where each program in the sequence may successively remove some of the remaining models. This mimics a network of leaders, where the first leader has the first say and may remove models that he or she finds unsatisfactory. Using this particular communication mechanism allows us to capture the entire polynomial hierarchy. This means, in particular, that communicating ASP could be used to solve problems that are above the second level of the polynomial hierarchy, such as some forms of abductive reasoning as well as PSPACE-complete problems such as STRIPS planning.
  \end{abstract}


%%%%%%%%%%%%%%%%%%%%%%%%%%%%%
%%%%%%%%%%%%%%%%%%%%%%%%%%%%%
%%%
%%%   INTRODUCTION
%%%
%%%%%%%%%%%%%%%%%%%%%%%%%%%%%
%%%%%%%%%%%%%%%%%%%%%%%%%%%%%

\section{Introduction}\label{sec:introduction}
Answer set programming (ASP) is a form of non-monotonic reasoning based on the stable model semantics~\cite{gelfond:stablemodel}. ASP has proven successful as an elegant and convenient vehicle for commonsense reasoning in discrete domains and to encode combinatorial optimization problems in a purely declarative way. It has been applied in, for example, plan generation~\cite{lifschitz:plan}, diagnosis~\cite{eiter:diagnosis} and biological networks~\cite{gebser:repair}. Being an active field of research, a large body of extensions have been proposed that improve upon the basics of ASP and offer, for example, cardinality constraints~\cite{niemela:extending} or nested expressions~\cite{lifschitz:nested}. Not all of these extensions provide an increase in terms of computational expressiveness and some are merely convenient syntactic sugar. 

One particularly interesting extension of ASP is called communicating ASP. It allows for a number of ASP programs to communicate, \ie share information about their knowledge base, giving them the ability to cooperate with each other to solve the problem at hand. Each ASP program involved, called a component program, has individual beliefs and reasoning capabilities. One of the benefits of this extension is that it eases the declarative formulation of a problem originating in a multi-agent context. Quite a number of different flavours of communicating ASP have been proposed in the literature, both in the narrow domain of ASP and in the broader domain of logic programming, where each of these papers presents intriguing examples that highlight the usefulness of communicating ASP, for example \cite{acqua:communicating}, \cite{brain:oclp}, \cite{roelofsen:minimal}, \cite{devos:laima} and \cite{nieuwenborgh:hierarchical}. In particular, all the examples involve multi-dimensional problems (\eg a police investigation with multiple witnesses) where each agent only has the knowledge contained in one or a few of the dimensions (\eg the witness only saw the burglar enter the building). 

\begin{figure}[ht]
  \includegraphics[scale=0.3]{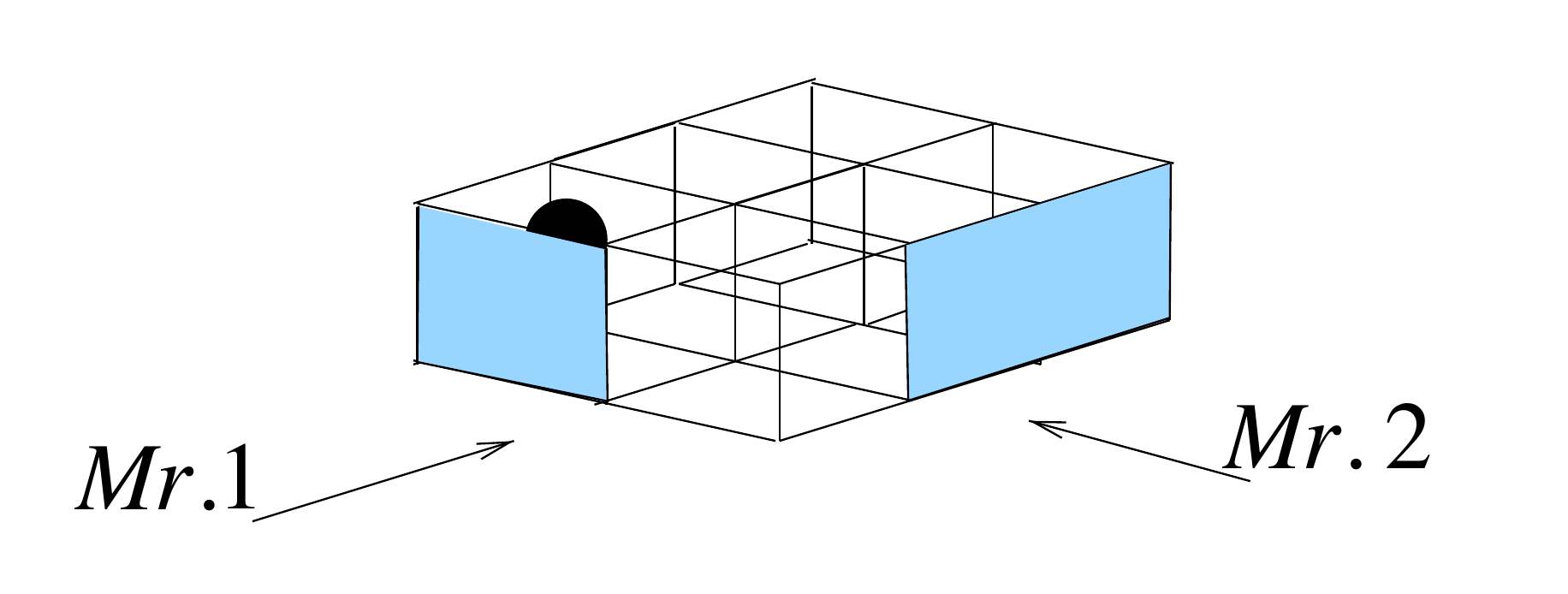}
  \caption{A magic box.}
  \label{fig:magicbox}
\end{figure}

A standard example to illustrate the usefulness in this area is shown in~\thefig{magicbox}.\footnote{Illustration from~\cite{roelofsen:minimal} used with permission from the authors.} The figure depicts two people who are looking at a box. The box is called magic because neither agent can make out its depth. The information the agents know is further limited because parts of the box are blinded. By cooperation, both agents can pinpoint the exact location of the ball. Indeed, $\mathit{Mr. 2}$ sees a ball on his left side. From this information $\mathit{Mr. 1}$ knows that there is a ball and that it must therefore be on his left (since he cannot see a ball on his right). This knowledge can be relayed back to $\mathit{Mr. 2}$. Both agents now know the exact position and depth of the ball.

Complexity results from~\cite{brewka:contextual} show that computing the answer sets of a communicating logic program is in \cNP.  In general, however, only few results exist regarding the expressiveness of such communicating ASP programs. In addition, for many of the known results, it is not clear whether an increase in expressiveness is due to the type of component programs considered (\ie the expressiveness of the individual programs involved) or due to the use of communication.

In communicating ASP, the notion of an answer set can be defined in different ways; we refer to the specific definition that is used as the communication mechanism. Answer set semantics is based on the idea of stable minimal models. When dealing with agents that can communicate, it becomes unclear how we should interpret the notion of minimality. One option is to assume global minimality, \ie we minimise over the conclusions of all the agents in the network. Another option is to assume minimality on the level of a single agent. Since in general it is not possible to find a model that is minimal for all individual agents, the order in which we minimise over the agents matters.

The main contributions of this paper are as follows. We present a systematic study of the additional expressiveness offered by allowing ASP programs to communicate, where, for the most part, we deliberately only consider component programs that are simple, \ie programs for which computing the answer sets is in~$\cP$. Clearly any step we move up in the polynomial hierarchy is then due to the ability of these simple ASP programs to communicate and collaborate  with each other. We show that the problem of deciding whether a literal is in any answer set of a communicating ASP program using simple communication is in \cNP. We also provide a simulation using a network of simple programs that is capable of modeling any program with negation-as-failure. These results are extended with  complexity results for when the network consists of disjunctive programs. In all cases, we examine the complexity of determining whether an answer set exists as well as whether a literal is true in any (brave reasoning) or all answer sets (cautious reasoning). Furthermore, we introduce the notion of multi-focused answer sets of communicating ASP programs, which allows us to successively focus (\ie minimise) on different agents. As it turns out, using multi-focused answer set programs it is possible to express any problem in PSPACE. This means in particular that communicating ASP could be used to solve problems that are above the second level of the polynomial hierarchy, such as some forms of abductive reasoning~\cite{eiter:complexity} as well as PSPACE-complete problems such as STRIPS planning~\cite{bylander:computational}. We show that these results hold for any choice of program in the network, being it either simple, normal or disjunctive, and we provide the complexity of determining whether a multi-focused answer set exists.

This paper aggregates and extends our work from~\cite{bauters:communicating,bauters:communicatingpolynomial}. The first work is extended with completeness results and an elaborate overview of the simulation of negation-as-failure. Furthermore, we provide more detailed complexity results including the complexity of determining whether an answer set exists and the complexity of determining whether some literal is true in some or all of the answer sets. The focused answer sets introduced in \thesec{focused} are from~\cite{bauters:communicatingpolynomial} and are more general than the corresponding notion that we introduced in~\cite{bauters:communicating}. Additional results are provided, such as the complexity of determining whether a multi-focused answer set exists, as well as new results concerning the use of disjunctive programs as component programs. We also provide the proofs of all results in the appendix.

%%%%%%%%%%%%%%%%%%%%%%%%%%%%%
%%%%%%%%%%%%%%%%%%%%%%%%%%%%%
%%%
%%%   PRELIMINARIES
%%%
%%%%%%%%%%%%%%%%%%%%%%%%%%%%%
%%%%%%%%%%%%%%%%%%%%%%%%%%%%%

\section{Background on Answer set programming}\label{sec:asp}
We first recall the basic concepts and results from ASP that are used in this paper. To define ASP programs, we start from a countable set of atoms and we define a~\concept{literal} $l$ as an atom $a$ or its classical negation $\neg a$. If $L$ is a set of literals, we use $\neg L$ to denote the set \condset{\neg l}{l \in L} where, by definition, $\neg\neg a = a$. A set of literals $L$ is \concept{consistent} if $L \cap \neg L = \emptyset$. An~\concept{extended literal} is either a literal or a literal preceded by $\snaf$ which we call the negation-as-failure operator. Intuitively we say that $\naf l$ is true when we have no proof to support $l$.
For a set of literals $L$, we use $\snaf(L)$ to denote the set \condset{\naf l}{l \in L}. 

A \concept{disjunctive rule} is an expression of the form \arule{\gamma}{(\alpha \cup \snaf(\beta))} where $\gamma$ is a set of literals (interpreted as a disjunction, denoted as $\srangec{l}{1}{n}{;}$) called the head of the rule and $(\alpha \cup \snaf(\beta))$ (interpreted as a conjunction) is the body of the rule with $\alpha$ and $\beta$ sets of literals. When the body is empty, the rule is called a~\concept{fact}. When the head is empty, the rule is called a~\concept{constraint}. In this paper, we do not consider constraints as they can readily be simulated using extended literals.\footnote{We can simulate a constraint $(\arule{}{body})$ by $(\arule{\mathit{fail}}{\naf \mathit{fail}, body)}$ with $\mathit{fail}$ a fresh atom.} A \concept{positive disjunctive rule} is a disjunctive rule without negation-as-failure in the body, \ie with $\beta = \emptyset$. A \concept{disjunctive program} $P$ is a finite set of disjunctive rules. The  \concept{Herbrand base} \hbase{P} of $P$ is the set of atoms appearing in program~$P$. A (partial) \concept{interpretation} $I$ of $P$ is any consistent set of literals $I \subseteq (\hbase{P} \cup \neg \hbase{P})$. $I$ is \concept{total} iff $I \cup \neg I = \hbase{P} \cup \neg \hbase{P}$.

A \concept{normal rule} is a disjunctive rule with exactly one literal $l$ in the head. A \concept{normal program} $P$ is a finite set of normal rules. A \concept{simple rule} is a normal rule without negation-as-failure in the body. A \concept{simple program} $P$ is a finite set of simple rules.  

The satisfaction relation $\models$ is defined for an interpretation $I$ as $I\models l$ iff $l \in I$, otherwise $I \not\models l$.
For an interpretation $I$ and $L$ a set of literals, we define $I \models L$ iff $\sForall{l \in L}{I \models l}$. An interpretation $I$ is a \concept{model} of a positive disjunctive rule  $r = \arule{\gamma}{\alpha}$, denoted $I \models r$, if $I \not\models \neg\gamma$ or $I \not\models \alpha$, \ie the body is false or at least one of the literals in the head can be true. 
An interpretation $I$ of a positive disjunctive program $P$ is a \concept{model} of $P$ iff $\sForall{r \in P}{I \models r}$.

Answer sets are defined using the \concept{immediate consequence operator} $T_P$ for a simple program $P$ \wrt an interpretation $I$ as
  \begin{equation}
    T_P(I) = I \cup \condset{l}{((\arule{l}{\alpha}) \in P)\wedge(\alpha \subseteq I)}.
  \end{equation}
  We use $P^{\star}$ to denote the fixpoint which is obtained by repeatedly applying $T_P$ starting from the empty interpretation, \ie the least fixpoint of $T_P$ \wrt set inclusion. An interpretation $I$ is an \concept{answer set} of a simple program $P$ iff $I = P^{\star}$. 
  
  The \concept{reduct $P^I$} of a disjunctive program $P$ \wrt the interpretation $I$ is defined as 
  $
  P^I = \condset{\arule{\gamma}{\alpha}}{(\arule{\gamma}{\alpha \cup \snaf(\beta)}) \in P, \beta \cap I = \emptyset}.
  $
We say that $I$ is an answer set of the disjunctive program $P$ when $I$ is a minimal model of $P^I$ \wrt set inclusion. 

 In the specific case of normal programs, answer sets can also be characterised in terms of fixpoints.  Specifically, it is easy to see that in this case the reduct $P^I$ is a simple program. We say that $I$ is an answer set of a normal program $P$ iff $\left({P^I}\right)^{\star} = I$, \ie if $I$ is the answer set of the reduct $P^{I}$. 

\section{Communicating programs}\label{sec:communicating}

The underlying intuition of communication between ASP programs is that of a function call or, in terms of agents, asking questions to other agents. This communication is based on a new kind of literal `$\ask{Q}{l}$', as in~\cite{giunchiglia:multilanguage,roelofsen:minimal,brewka:equilibria}. 
If the literal $l$ is not in the answer set of program $Q$ then \ask{Q}{l} is false; otherwise \ask{Q}{l} is true. The semantics presented in this section are closely related to the minimal semantics of \cite{brewka:equilibria} and especially the semantics of \cite{buccafurri:logic}.

  Let \struprog{P} be a finite set of program names. A \concept{\struprog{P}-situated literal} is an expression of the form \ask{Q}{l} with $Q \in \struprog{P}$ and $l$ a literal. For $R \in \struprog{P}$, a \struprog{P}-situated literal $\ask{Q}{l}$ is called \concept{$R$-local} if $Q=R$. For a set of literals $L$, we use $\ask{Q}{L}$ as a shorthand for $\condset{\ask{Q}{l}}{l\in L}$. For a set of \struprog{P}-situated literals $X$ and $Q\in\struprog{P}$, we use $\proj{X}{Q}$ to denote $\condset{l}{\ask{Q}{l}\in X}$, i.e. the projection of $X$ on $Q$. A set of \struprog{P}-situated literals $X$ is \concept{consistent} iff $\proj{X}{Q}$ is consistent for all $Q\in\struprog{P}$. By $\neg X$ we denote the set \condset{\ask{Q}{\neg l}}{\ask{Q}{l} \in X} where we define $\ask{Q}{\neg\neg l} = \ask{Q}{l}$. An \concept{extended \struprog{P}-situated literal} is either a \struprog{P}-situated literal or a \struprog{P}-situated literal preceded by $\snaf$. For a set of \struprog{P}-situated literals $X$, we use $not(X)$ to denote the set \condset{\naf \ask{Q}{l}}{\ask{Q}{l} \in X}. For a set of extended \struprog{P}-situated literals $X$ we denote by $X_{\mathrm{pos}}$ the set of \struprog{P}-situated literals in $X$, \ie those extended \struprog{P}-situated literals in $X$ that are not preceded by negation-as-failure, while $X_{\mathrm{neg}} = \condset{\ask{Q}{l}}{\naf \ask{Q}{l} \in X}$. 
  
  A \concept{\struprog{P}-situated disjunctive rule} is an expression of the form $\arule{\ask{Q}{\gamma}}{(\alpha \cup not(\beta))}$ where $\gamma$ is a set of literals, called the head of the rule, and $(\alpha \cup \snaf(\beta))$ is called the body of the rule with $\alpha$ and $\beta$ sets of $\struprog{P}$-situated literals. A \struprog{P}-situated disjunctive rule \arule{\ask{Q}{\gamma}}{(\alpha \cup \naf(\beta))} is called $R$-local whenever $Q=R$. A \concept{\struprog{P}-component disjunctive program} $Q$ is a finite set of $Q$-local $\struprog{P}$-situated disjunctive rules. Henceforth we shall use $\struprog{P}$ both to denote the set of program names and to denote the set of actual \struprog{P}-component disjunctive programs. A \concept{communicating disjunctive program} \struprog{P} is then a finite set of $\struprog{P}$-component disjunctive programs.  
  
  A \concept{\struprog{P}-situated normal rule} is an expression of the form $\arule{\ask{Q}{l}}{(\alpha \cup not(\beta))}$ where $\ask{Q}{l}$ is a single \struprog{P}-situated literal. A \concept{\struprog{P}-situated simple rule} is an expression of the form \arule{\ask{Q}{l}}{\alpha}, \ie a \struprog{P}-situated normal rule without negation-as-failure. A \concept{\struprog{P}-component normal (resp. simple) program} $Q$ is a finite set of $Q$-local $\struprog{P}$-situated normal (resp. simple) rules. A \concept{communicating normal (resp. simple) program} \struprog{P} is then a finite set of $\struprog{P}$-component normal (resp. simple) programs.
  
  In the remainder of this paper we drop the $\struprog{P}$-prefix whenever the set $\struprog{P}$ is clear from the context. Whenever the name of the component disjunctive program $Q$ is clear, we write $l$ instead of $\ask{Q}{l}$ for $Q$-local situated literals. Note that a communicating disjunctive (resp. normal, simple) program with only one component program thus trivially corresponds to a classical disjunctive (resp. normal, simple) program. Finally, for notational convenience, we write communicating program when it is clear from the context whether the program is a communicating simple program or a communicating normal program. 

\begin{example}\label{ex:terminology}
Consider the communicating normal program $\struprog{P} = \set{Q,R}$ with the following situated rules:
\begin{align*}
  \arule{\ask{Q}{a}}{\ask{R}{a}}&&\arule{\ask{Q}{b}&}{}&&\arule{\ask{Q}{c}}{\ask{Q}{c}}\\
  \arule{\ask{R}{a}}{\ask{Q}{a}}&&\arule{\ask{R}{b}&}{\naf\ask{Q}{c}}.&&
\end{align*}
$\ask{Q}{a}$, $\ask{Q}{b}$, $\ask{Q}{c}$, $\ask{R}{a}$ and $\ask{R}{b}$ are situated literals. The situated simple rules on the top line are $Q$-local since we respectively have $\ask{Q}{a}$, $\ask{Q}{b}$ and  $\ask{Q}{c}$ in the head of these rules. The situated normal rules on the bottom line are $R$-local. Hence $Q = \set{\arule{a}{\ask{R}{a}}, \arule{b}{}, \arule{c}{c}}$ and $R = \set{\arule{a}{\ask{Q}{a}}, \arule{b}{\naf \ask{Q}{c}}}$.
\end{example}

Similar as for a classical program, we can define the \concept{Herbrand base} for a component program $Q$ as the set of atoms $\hbase{Q}$ = \condset{a}{\ask{Q}{a} \text{ or } \ask{Q}{\neg a} \text{ appearing in } Q}, \ie the set of atoms occurring in the $Q$-local situated literals in $Q$. We then define $\hbase{\struprog{P}} = \condset{\ask{Q}{a}}{Q \in \struprog{P} \mbox{ and } a \in \bigcup_{R \in \struprog{P}}\hbase{R}}$ as the Herbrand base of the communicating program \struprog{P}.

\begin{example}\label{ex:herbrand}
Given the communicating normal program $\struprog{P} = \set{Q,R}$ from \themyex{terminology} we have that $\hbase{Q} = \set{a,b,c}$, $\hbase{R} = \set{a,b}$ and $\hbase{\struprog{P}} = \set{\ask{Q}{a},\ask{Q}{b}, \ask{Q}{c},\ask{R}{a},\ask{R}{b}, \ask{R}{c}}$.
\end{example}

 We say that a (partial) interpretation $I$ of a communicating disjunctive program $\struprog{P}$ is any consistent subset $I \subseteq \left(\hbase{\struprog{P}} \cup \neg\hbase{\struprog{P}}\right)$. Given an interpretation $I$ of a communicating disjunctive program $\struprog{P}$, the reduct $Q^I$ for $Q \in \struprog{P}$ is the component disjunctive program obtained by deleting
  \begin{itemize}
    \item each rule with an extended situated literal `$\naf \ask{R}{l}$' such that $\ask{R}{l} \in I$;
    \item each remaining extended situated literal of the form `$\naf \ask{R}{l}$';
    \item each rule with a situated literal `$\ask{R}{l}$' that is not $Q$-local such that $\ask{R}{l} \notin I$;
    \item each remaining situated literal `$\ask{R}{l}$' that is not $Q$-local.% and such that $\ask{R}{l} \in I$.
  \end{itemize}

Note that this definition actually combines two types of reducts together. On the one hand, we remove the negation-as-failure according to the given knowledge. On the other hand, we also remove situated literals that are not $Q$-local, again according to the given knowledge. The underlying intuition of the reduct remains unchanged compared to the classical case: we take the information into account which is encoded in the guess $I$ and we simplify the program so that we can easily verify whether or not $I$ is stable, \ie whether or not $I$ is a minimal model of the reduct. Analogous to the definition of the reduct for disjunctive programs~\cite{gelfond:classical}, the reduct of a communicating disjunctive program thus defines a way to reduce a program relative to some guess $I$. The reduct of a communicating disjunctive program is a communicating disjunctive program (without negation-as-failure) that only contains component disjunctive programs $Q$ with $Q$-local situated literals. That is, each remaining component disjunctive program $Q$ corresponds to a classical disjunctive program.

\begin{example}
Let us once again consider the communicating normal program $\struprog{P} = \set{Q,R}$ from \themyex{terminology}. Given $I = \set{\ask{Q}{a}, \ask{Q}{b}, \ask{R}{a}, \ask{R}{b}}$ we find that $Q^I = \set{\arule{a}{}, \arule{b}{}, \arule{c}{c}}$ and $R^I = \set{\arule{a}{}, \arule{b}{}}$. We can easily treat $Q^I$ and $R^I$ separately since they now correspond to classical programs.
\end{example}

\begin{definition}\label{def:answerset-communicating}
  We say that an interpretation $I$ of a communicating disjunctive program $\struprog{P}$ is an \concept{answer set} of \struprog{P} if and only if $\sForall{Q \in \struprog{P}}{(\ask{Q}{\proj{I}{Q}})}$ is the minimal model \wrt set inclusion of $Q^I$. In other words: an interpretation $I$ is an answer set of a communicating disjunctive program $\struprog{P}$ if and only if for every component program $Q$ we have that the projection of $I$ on $Q$ is an answer set of the component program $Q^I$ under the classical definition. 
\end{definition}

In the specific case of a communicating normal program $\struprog{P}$ we can equivalently say that $I$ is an answer set of \struprog{P} if and only if we have that $\sForall{Q \in \struprog{P}}{(\ask{Q}{\proj{I}{Q}}) = \left({Q^I}\right)^\star}$.

\begin{example}
The communicating normal program $\struprog{P} = \set{Q,R}$ from \themyex{terminology} has two answer sets, namely $\set{\ask{Q}{b}, \ask{R}{b}}$ and $\set{\ask{Q}{a}, \ask{Q}{b}, \ask{R}{a}, \ask{R}{b}}$.
\end{example}

Note that while most approaches do not allow self-references of the form \arule{\ask{Q}{a}}{\ask{Q}{a}}, in our approach this poses no problems as it is semantically equivalent to \arule{\ask{Q}{a}}{a}. Also note that our semantics allow for ``mutual influence'' as in \cite{brewka:equilibria,buccafurri:logic} where the belief of an agent can be supported by the agent itself, via belief in other agents, \eg \set{\arule{\ask{Q}{a}}{\ask{R}{a}},\arule{\ask{R}{a}}{\ask{Q}{a}}}. Furthermore we want to point out that the belief between agents is the belief as identified in~\cite{lifschitz:nested}, \ie the situated literal $\ask{Q}{l}$ is true in our approach whenever ``$\neg \naf \ask{Q}{l}$'' is true in the approach introduced in~~\cite{lifschitz:nested} for nested logic programs and treating $\ask{Q}{l}$ as a fresh atom.

Before we introduce our first proposition, we generalise the immediate consequence operator for (classical) normal programs to the case of communicating simple programs. Specifically, the operator $T_{\struprog{P}}$ is defined \wrt an interpretation $I$ of $\struprog{P}$ as 
$$T_{\struprog{P}}(I) = I \cup \condset{\ask{Q}{l}}{(\arule{l}{\alpha}) \in Q, Q \in \struprog{P}, \alpha \subseteq I}$$
where $\alpha$ is a set of $\struprog{P}$-situated literals. It is easy to see that this operator is monotone. Together with a result from \cite{tarski:lattice} we know that this operator has a least fixpoint. We use $\struprog{P}^{\star}$ to denote this fixpoint obtained by repeatedly applying $T_{\struprog{P}}$ starting from the empty interpretation. Clearly, this fixpoint can be computed in polynomial time. Furthermore, just like the immediate consequence operator for (classical) normal programs, this generalised operator only derives the information that is absolutely necessary, \ie the fixpoint $\struprog{P}^{\star}$ is globally minimal.

\begin{proposition}
\label{prop:one-simple-is-polynomial}
Let \struprog{P} be a communicating simple program. We then have that:
\begin{list}{\labelitemi}{\leftmargin=2em}
  \item there always exists at least one answer set of $\struprog{P}$;
  \item there is always a unique answer set of $\struprog{P}$ that is globally minimal;
  \item we can compute this unique globally minimal answer set in polynomial time. 
\end{list}
\end{proposition}

\begin{example}
Consider the communicating simple program $\struprog{P}$ with the rules
\begin{align*}
  \arule{\ask{Q}{a}}{\ask{R}{a}}&&\arule{\ask{R}{a}}{\ask{Q}{a}}&&\arule{\ask{Q}{b}}{}.
\end{align*}
This communicating simple program has two answer sets, namely $\set{\ask{Q}{a}, \ask{Q}{b}, \ask{R}{a}}$ and $\set{\ask{Q}{b}}$. We have that $\struprog{P}^{\star} = \set{\ask{Q}{b}}$, \ie $\set{\ask{Q}{b}}$ is the answer set that can be computed in polynomial time. Intuitively, this is the answer set of the communicating simple program $\struprog{P}$ where we treat every situated literal as an ordinary literal. For example, if we replace the situated literal $\ask{Q}{a}$ (resp. \ask{Q}{b}, \ask{R}{a}) by the literals $qa$ (resp. $qb, ra$) we obtain the simple program 
\begin{align*}
\arule{qa}{ra} && \arule{qb}{}&& \arule{ra}{qa}
\end{align*}
which has the unique answer set $\set{qb}$, with $qb$ the literal that replaced $\ask{Q}{b}$.
Note that the procedure involving the generalised fixpoint does not allow us to derive the second answer set.  In general, no polynomial procedure will be able to verify whether there is some answer set in which a given literal is true (unless \cP=\cNP).
\end{example}

Although finding an answer set of a communicating simple program can be done in polynomial time, we will see in the next section that brave reasoning (the problem of determining whether a given situated literal $\ask{Q}{l}$ occurs in any answer set of a communicating simple program) is \cNP-hard. Consequently, cautious reasoning (the problem of determining whether a given literal $\ask{Q}{l}$ occurs in all answer sets of a communicating simple program) is $\ccoNP$-hard.

%\begin{example}\label{ex:communicating-answersets}
%Consider the communicating program $\struprog{P}_{intro}$ from Example~\ref{ex:intro}. It is easy to see that $M = \set{\ask{H_1}{\neg a}, \ask{H_2}{a}, \ask{H_2}{p}, \ask{U}{\neg o}, \ask{U}{c}}$ is the unique answer set of $\struprog{P}_{intro}$. Indeed, we obtain the reducts ${(H_1)}^{M} = \set{\arule{\neg a}{}}$, ${(H_2)}^{M} = \set{\arule{a}{}, \arule{p}{}}$ and ${(U)}^{M} = \set{\arule{\neg o}{}, \arule{c}{}}$ which have the answer sets $\set{\neg a}, \set{a, p}$ and $\set{\neg o, c}$, respectively.
%\end{example}

%%%%%%%%%%%%%%%%%%%%%%%%%%%%%
%%%%%%%%%%%%%%%%%%%%%%%%%%%%%
%%%
%%%   SIMULATING NEGATION-AS-FAILURE WITH COMMUNICATION
%%%
%%%%%%%%%%%%%%%%%%%%%%%%%%%%%
%%%%%%%%%%%%%%%%%%%%%%%%%%%%%

\section{Simulating Negation-as-Failure with Communication}\label{sec:simulating}
The addition of communication to ASP programs can provide added expressiveness over simple programs and a resulting increase in computational complexity for brave reasoning and cautious reasoning. To illustrate this observation, in this section we show that a communicating simple program can simulate normal programs.\footnote{Recall that simple programs are \cP-complete and normal programs are \cNP-complete~\cite{baral:knowledge}.} Furthermore, we illustrate that, surprisingly, there is no difference in terms of computational complexity between communicating simple programs and communicating normal programs; a communicating simple program can be constructed which simulates any given communicating normal program.

For starters, we recall some of the notions of complexity theory. The complexity classes \spoldelta{n}, \spolsig{n} and \spolpi{n} are defined as follows, for $i \in \natural$~\cite{papadimitriou:computational}:
\begin{align*}
  \spoldelta{0} &= \spolsig{0} = \spolpi{0} = \cP\\
  \spoldelta{i+1} &= \cP^{\spolsig{i}}\\
  \spolsig{i+1} &= \cNP^{\spolsig{i}}\\
  \spolpi{i+1} &= \cco\left(\spolsig{i+1}\right)
\end{align*}
where $\cNP^{\spolsig{i}}$ (resp. $\cP^{\spolsig{i}}$) is the class of problems that can be solved in polynomial time on a non-deterministic machine (resp. deterministic machine) with an \spolsig{i} oracle, \ie assuming a procedure that can solve \spolsig{i} problems in constant time. 
For a general complexity class $\cC$, a problem is $\cC$-hard if any other problem in \cC\ can be efficiently reduced to this problem %(\ie in polynomial time for problems in $\cNP$ or higher). 
A problem is said to be \cC-complete if the problem is in \cC\ and the problem is \cC-hard.
 Deciding the validity of a QBF $\phi = \exists X_1 \forall X_2 ... \mathrm{\Theta} X_n \cdot p(X_1, X_2, \cdots X_n)$ with $\Theta = \exists$ if $n$ is odd and $\Theta = \forall$ otherwise, is the canonical $\spolsig{n}$-complete problem. Deciding the validity of a QBF $\phi = \forall X_1 \exists X_2 ... \mathrm{\Theta} X_n \cdot p(X_1, X_2, \cdots X_n)$ with $\Theta = \forall$ if $n$ is odd and $\Theta = \exists$ otherwise, is the canonical $\spolpi{n}$-complete problem. Moreover, these results also hold when we restrict ourselves to problems with $p(X_1, X_2, \cdots X_n)$ in disjunctive normal form, except when the last quantifier is an $\exists$.\footnote{Given a QBF with the last quantifier an $\exists$ and a formula in disjunctive normal form, we can reduce the problem in polynomial time to a new QBF without the last quantifier. To do this, for every variable quantified by this last quantifier we remove those clauses in which both the quantified variable and its negation occur (contradiction) and then remove all occurrences of the quantified variables in the remaining clauses as well as the quantifier itself. The new QBF is then valid if and only if the original QBF is valid.} Brave reasoning as well as answer set existence for simple, normal and disjunctive programs is $\cP$-complete, $\cNP$-complete and $\spolsig{2}$-complete, respectively~\cite{baral:knowledge}. Cautious reasoning for simple, normal and disjunctive programs is $\cco\cP$-complete, $\ccoNP$-complete and $\cco\spolsig{2}$-complete~\cite{baral:knowledge}.

In this section we start by giving an example of the transformation that allows us to simulate (communicating) normal programs using communicating simple programs. A formal definition of the simulation is given below in~\themydef{simulateextended}. The correctness is proven by Propositions~\ref{prop:partialanswerset} and \ref{prop:converse}. 

\begin{example}\label{ex:simulation}
  Consider the communicating normal program \struprog{P} with the rules
  \begin{align*}
    \arule{\ask{Q_1}{a}&}{\naf \ask{Q_2}{b}}\\
    \arule{\ask{Q_2}{b}&}{\naf \ask{Q_1}{a}}.
  \end{align*}
  Note that if we were to take $Q_1 = Q_2$ then this example corresponds to a normal program. In our simulation, the communicating normal program \struprog{P} is transformed into the following communicating simple program $\struprog{P'} = \set{Q'_1, Q'_2, N_1, N_2}$:
  \begin{align*}
    \arule{\ask{Q'_1}{a}&}{\ask{N_2}{\neg\fresh{ b}}}&\arule{\ask{N_1}{\fresh{a}}&}{\ask{Q'_1}{a}}\\
    \arule{\ask{Q'_2}{b}&}{\ask{N_1}{\neg\fresh{ a}}}&\arule{\ask{N_2}{\fresh{b}}&}{\ask{Q'_2}{b}}\\
    \arule{\ask{Q'_1}{\neg\fresh{ a}}&}{\ask{N_1}{\neg\fresh{ a}}}&\arule{\ask{N_1}{\neg\fresh{ a}}&}{\ask{Q'_1}{\neg\fresh{ a}}}\\
    \arule{\ask{Q'_2}{\neg\fresh{ b}}&}{\ask{N_2}{\neg\fresh{ b}}}&\arule{\ask{N_2}{\neg\fresh{ b}}&}{\ask{Q'_2}{\neg\fresh{ b}}}.
  \end{align*}
  The transformation creates two types of component programs or \emph{`worlds'}, namely $Q'_i$ and $N_i$. The component program $Q'_i$ is similar to $Q_i$ but occurrences of extended situated literals of the form $\naf \ask{Q_i}{l}$ are replaced by $\ask{N_i}{\neg \fresh{l}}$, with $\fresh{l}$ a fresh literal. The non-monotonicity associated with negation-as-failure is simulated by introducing the rules $\arule{\neg \fresh{l}}{\ask{N_i}{\neg\fresh{ l}}}$ and $\arule{\neg\fresh{ l}}{\ask{Q'_i}{\neg\fresh{ l}}}$ in $Q'_i$ and $N_i$, respectively. Finally, we add rules of the form $\arule{\fresh{l}}{\ask{Q'_i}{l}}$ to $N_i$, creating an inconsistency when $N_i$ believes $\neg\fresh{ l}$ and $Q'_i$ believes $l$.
  
The resulting communicating simple program \struprog{P'} is an equivalent program in that its answer sets correspond to those of the original communicating normal program, yet without using negation-as-failure. Indeed, the answer sets of \struprog{P} are $\set{\ask{Q_1}{a}}$ and $\set{\ask{Q_2}{b}}$ and the answer sets of \struprog{P'} are $\set{\ask{Q'_1}{a}} \cup \set{\ask{Q'_2}{\neg\fresh{ b}}, \ask{N_2}{\neg\fresh{ b}}, \ask{N_1}{\fresh{a}}}$ and $\set{\ask{Q'_2}{b}} \cup \set{\ask{Q'_1}{\neg\fresh{ a}}, \ask{N_1}{\neg\fresh{ a}}, \ask{N_2}{\fresh{b}}}$. 
\end{example}

Note that the simulation given in \themyex{simulation} can in fact be simplified. Indeed, in this particular example there is no need to have two additional component programs $N_1$ and $N_2$ since $Q_1$ and $Q_2$ do not share literals. Also, in this particular example, we need not use `$\fresh{a}$' and `$\fresh{b}$' since the simulation would work just as well if we simply considered `$a$' and `$b$' instead. Nonetheless, for the generality of the simulation such technicalities are necessary. Without adding an additional component program $N_i$ for every original component program $Q_i$ the simulation would in general not work when two component programs shared literals, \eg $\ask{Q_1}{a}$ and $\ask{Q_2}{a}$. Furthermore, we need to introduce fresh literals as otherwise the simulation would in general not work when we had true negation in the original program, \eg $\ask{Q}{\neg a}$. We now give the definition of the simulation which works in the general case.

\begin{definition}\label{def:simulateextended}
  Let $\struprog{P} = \set{\srange{Q}{1}{n}}$ be a communicating normal program. The communicating simple program $\struprog{P'} = \set{\srange{Q'}{1}{n}, \srange{N}{1}{n}}$ with $1 \leq i,j \leq n$ that simulates \struprog{P} is defined by
  \begin{eqnarray}
    Q'_i&=&\condset{\arule{l}{\alpha'_{\mathrm{pos}} \cup \condset{\ask{N_j}{\neg\fresh{ k}}}{\ask{Q_j}{k} \in \alpha_{\mathrm{neg}} }}}{(\arule{l}{\alpha}) \in Q_i}\label{simulation-Q-pos}\\
    &\cup& \condset{\arule{\neg\fresh{ b}}{\ask{N_i}{\neg\fresh{ b}}}}{\ask{Q_i}{b} \in \struprog{E}_{\mathrm{neg}}}\label{simulation-Q-neg}\\
    N_i&=&\condset{\arule{\neg\fresh{ b}}{\ask{Q'_i}{\neg\fresh{ b}}}}{\ask{Q_i}{b} \in \struprog{E}_{\mathrm{neg}}}\label{simulation-N}\\
    &\cup& \condset{\arule{\fresh{b}}{\ask{Q'_i}{b}}}{\ask{Q_i}{b} \in \struprog{E}_{\mathrm{neg}}}\label{constraint-N}
  \end{eqnarray}
  with $\alpha' = \condset{\ask{Q'_{j}}{l}}{\ask{Q_{j}}{l} \in \alpha}$, $
    \struprog{E}_{\mathrm{neg}} = \bigcup^n_{i=1}\left(\bigcup_{\left(\arule{a}{\alpha}\right) \in Q_i} \alpha_{\mathrm{neg}}\right)$ and with $\alpha_{\mathrm{pos}}$ and $\alpha_{\mathrm{neg}}$ as defined before. Note how this is a polynomial transformation with at most $3 \cdot \card{\struprog{E}_{\mathrm{neg}}}$ additional rules. This is important when later we use the \cNP-completeness results from normal programs to show that communicating simple programs are \cNP-complete as well. Recall that both $\neg\fresh{ b}$ and $\fresh{b}$ are fresh literals that intuitively correspond to $\neg b$ and $b$. We use ${Q'_i}+$ to denote the set of rules in $Q'_i$ defined by \Ref{simulation-Q-pos} and ${Q'_i}-$ to denote the set of rules in $Q'_i$ defined by \Ref{simulation-Q-neg}. 
\end{definition}

The intuition of the simulation in \themydef{simulateextended} is as follows. The simulation uses the property of mutual influence to mimic the choice induced by negation-as-failure. This is obtained from the interplay between rules \eqref{simulation-Q-neg} and \eqref{simulation-N}. As such, we can use the new literal `$\neg\fresh{b}$' instead of the original extended (situated) literal `$\naf b$', allowing us to rewrite the rules as we do in \eqref{simulation-Q-pos}. In order to ensure that the simulation works even when the program we want to simulate already contains classical negation, we need to specify some additional bookkeeping~\eqref{constraint-N}.

As will become clear from \themyprop{partialanswerset} and \themyprop{converse}, the above transformation preserves the semantics of the original program. Since we can thus rewrite any normal program as a communicating normal program, the importance is twofold. On one hand, we reveal that communicating normal programs do not have any additional expressive power over communicating simple programs. On the other hand, it follows that communicating simple programs allow us to solve $\cNP$-complete problems. Before we show the correctness of the simulation in \themydef{simulateextended}, we introduce a lemma.

\begin{lemma}\label{lem:analogousanswerset}
Let $\struprog{P} = \set{\srange{Q}{1}{n}}$ and let $\struprog{P}' = \set{\srange{Q'}{1}{n}, \srange{N}{1}{n}}$ with $\struprog{P}$ a communicating normal program and $\struprog{P}'$ the communicating simple program that simulates \struprog{P} defined in \themydef{simulateextended}. Let $M$ be an answer set of \struprog{P} and let the interpretation $M'$ be defined~as:
  \begin{align}
  \begin{split}
    M' = &\condset{\ask{Q'_i}{a}}{\ask{Q_i}{a} \in M}\\
    &\cup~\condset{\ask{Q'_i}{\neg\fresh{ b}}}{\ask{Q_i}{b} \notin M}\\
    &\cup~\condset{\ask{N_i}{\neg\fresh{ b}}}{\ask{Q_i}{b} \in M}\\
    &\cup~\condset{\ask{N_i}{\fresh{a}}}{\ask{Q_i}{a} \in M}.
    \end{split}
  \end{align}
%  \begin{align}
%  \begin{split}
%    M' = &\condset{\ask{Q'_i}{a}}{a \in \proj{M}{Q_i}, Q_i \in \struprog{P}}\\
%    &\cup~\condset{\ask{Q'_i}{\neg\fresh{ b}}}{b \notin \proj{M}{Q_i}, Q_i \in \struprog{P}}\\
%    &\cup~\condset{\ask{N_i}{\neg\fresh{ b}}}{b \notin \proj{M}{Q_i}, Q_i \in \struprog{P}}\\
%    &\cup~\condset{\ask{N_i}{\fresh{a}}}{a \in \proj{M}{Q_i}, Q_i \in \struprog{P}}.
%    \end{split}
%  \end{align}
  For each $i \in \set{\range{1}{n}}$ it holds that $(Q'_i+)^{M'} = \condset{\arule{l}{\alpha'}}{\arule{l}{\alpha} \in Q^M_i}$ with $Q'_i+$ the set of rules defined in \eqref{simulation-Q-pos} with $\alpha' = \condset{\ask{Q'_{i}}{b}}{\ask{Q_{i}}{b} \in \alpha}$.
\end{lemma}

Using this lemma, we can prove that $M'$ as defined in~\themylem{analogousanswerset} is indeed an answer set of the communicating simple program that simulates the communicating normal program $\struprog{P}$ when $M$ is an answer set of $\struprog{P}$. 

\begin{proposition}\label{prop:partialanswerset}
Let $\struprog{P} = \set{\srange{Q}{1}{n}}$ and let $\struprog{P}' = \set{\srange{Q'}{1}{n}, \srange{N}{1}{n}}$ with $\struprog{P}$ a communicating normal program and $\struprog{P}'$ the communicating simple program that simulates \struprog{P} as defined in \themydef{simulateextended}. If $M$ is an answer set of \struprog{P}, then $M'$ is an answer set of $\struprog{P}'$ with $M'$ defined as in \themylem{analogousanswerset}.
\end{proposition}

Next we introduce \themylem{reverseanalogousanswerset}, which is similar to \themylem{analogousanswerset} in approach but which states the converse.

%\begin{lemma}\label{lem:pkiffnk}
%Let $\struprog{P}' = \set{\srange{Q'}{1}{n}, \srange{N}{1}{n}}$ be a communicating simple program that simulates a communicating program $\struprog{P} = \set{\srange{Q}{1}{n}}$ as defined in \themydef{simulateextended}. If $M'$ is an answer set of $\struprog{P}'$, we must have that $\left(\ask{Q'_k}{\neg\fresh{ c}}\right) \in M'$ if and only if $\left(\ask{N_k}{\neg\fresh{ c}}\right) \in M'$ for any $k$ where $1 \leq k \leq n$.
%\end{lemma}

\begin{lemma}\label{lem:reverseanalogousanswerset}
Let $\struprog{P} = \set{\srange{Q}{1}{n}}$ and let $\struprog{P}' = \set{\srange{Q'}{1}{n}, \srange{N}{1}{n}}$ with $\struprog{P}$ a communicating normal program and $\struprog{P'}$ the communicating simple program that simulates \struprog{P}. Assume that $M'$ is an answer set of $\struprog{P'}$ and that $\proj{(M')}{N_i}$ is total \wrt $\hbase{N_i}$ for all $i \in \set{\range{1}{n}}$. Let $M$ be defined as
  \begin{align}
    M = & \condset{\ask{Q_i}{b}}{\ask{Q'_i}{b} \in \left(\left(Q'_i+\right)^{M'}\right)^{\star}}
  \end{align}
  For each $i \in \set{\range{1}{n}}$, it holds that $(Q'_i+)^{M'} = \condset{\arule{l}{\alpha'}}{\arule{l}{\alpha} \in Q^M_i}$ with $\alpha' = \condset{\ask{Q'_{i}}{b}}{\ask{Q_{i}}{b} \in \alpha}$.
\end{lemma}

\begin{proposition}\label{prop:converse}
Let $\struprog{P} = \set{\srange{Q}{1}{n}}$ and let $\struprog{P}' = \set{\srange{Q'}{1}{n}, \srange{N}{1}{n}}$ with $\struprog{P}$ a communicating normal program and $\struprog{P'}$ the communicating simple program that simulates \struprog{P}. Assume that $M'$ is an answer set of $\struprog{P'}$ and that $\proj{(M')}{N_i}$ is total \wrt $\hbase{N_i}$ for all $i \in \set{\range{1}{n}}$. Then the interpretation $M$ defined in \themylem{reverseanalogousanswerset} is an answer set of \struprog{P}. 
\end{proposition}

It is important to note that \themylem{reverseanalogousanswerset} and, by consequence, \themyprop{converse} require (part of) the answer set $M'$ to be total. This is a necessary requirement, as demonstrated by the following example. 

\begin{example}\label{ex:needs-total}
  Consider the normal program $R = \set{\arule{a}{\naf a}}$ which has no answer sets. The corresponding communicating simple program $\struprog{P}' = \set{Q',N}$ has the following rules:
  \begin{align*}
    \arule{\ask{Q'}{a}&}{\ask{N}{\neg\fresh{ a}}}&    \arule{\ask{N}{\neg\fresh{ a}}&}{\ask{Q'}{\neg\fresh{ a}}}\\
    \arule{\ask{Q'}{\neg\fresh{ a}}&}{\ask{N}{\neg\fresh{ a}}}&    \arule{\ask{N}{\fresh{a}}&}{\ask{Q'}{a}}.
  \end{align*}
It is easy to see that $I = \emptyset$ is an answer set of $\struprog{P}'$ since we have ${Q'}^I=N^I=\emptyset$. Notice that $I$ does not correspond with an answer set of $R$, which is due to $\proj{I}{N} = \emptyset$ not being total and hence we cannot apply~\themyprop{converse}.
\end{example}

Regardless, it is easy to see that the requirement for the answer set to be total can be built into the simulation program. Indeed, it suffices to introduce additional rules to every $N_i$ with $1 \leq i \leq n$ in the simulation defined in \themydef{simulateextended}. These rules are 
\begin{align*}
&\condset{\arule{\ask{N_i}{a}}{\ask{N_i}{\fresh{a}}},\arule{\ask{N_i}{a}}{\ask{N_i}{\neg\fresh{a}}}}{\fresh{a} \in \hbase{N_i}}\\&\cup \set{\arule{\ask{N_i}{total}}{\beta}} \text{ with } \beta =  \condset{\ask{N_i}{a}}{\fresh{a} \in \hbase{N_i}}.
\end{align*}
Thus the requirement that (part of) the answer set must be total can be replaced by the requirement that the situated literals $\ask{N_i}{total}$ must be true in the answer set. Hence, if we want to check whether a literal $\ask{Q}{l}$ is true in at least one answer set of a (communicating) normal program, it suffices to check whether $\ask{Q}{l}$ and $\ask{N_i}{total}$ can be derived in the communicating simple program that simulates it. Clearly we find that brave reasoning for communicating simple programs is \cNP-hard. 

What we have done so far is important for two reasons. First, we have shown that the complexity of brave reasoning for communicating normal programs is no harder than brave reasoning for communicating simple programs. Indeed, the problem of brave reasoning for communicating normal programs can be reduced in polynomial time to the problem of brave reasoning for communicating simple programs. Second, since normal programs are a special case of communicating normal programs and since we know that brave reasoning for normal programs is an \cNP-complete problem, we have successfully shown that brave reasoning for communicating simple programs is \cNP-hard. In order to show that brave reasoning for communicating simple programs is \cNP-complete, we need to additionally show that this is a problem in \cNP. To this end, consider the following algorithm to find the answer sets of a communicating simple program $\struprog{P}$:
\begin{align*}
\begin{minipage}{0.8\linewidth}
\textsf{guess an interpretation  $I \subseteq (\hbase{\struprog{P}} \cup \neg\hbase{\struprog{P}})$}\\
\textsf{verify that this interpretation is an answer set as follows:}\\
${}$\hspace{0.5cm}\textsf{calculate the reduct $Q^I$ of each component program $Q$}\\
${}$\hspace{0.5cm}\textsf{calculate the fixpoint of each simple component program $Q^I$}\\
${}$\hspace{0.5cm}\textsf{verify that $Q:\proj{I}{Q} = {(Q^I)}^{\star}$ for each component program $Q$}
\end{minipage}
\end{align*}

The first step of the algorithm requires a choice, hence the algorithm is non-deterministic. Next we determine whether this guess is indeed a communicating answer set, which involves taking the reduct, computing the fixpoint and verifying whether this fixpoint coincides with our guess. Clearly, verifying whether the interpretation is an answer set can be done in polynomial time and thus the algorithm to compute the answer sets of a communicating simple program is in \cNP, and thus \cNP-complete, regardless of the number of component programs. These same results hold for communicating normal programs since the reduct also removes all occurrences of negation-as-failure. 

For communicating disjunctive programs it is easy to see that the $\spolsig{2}$-completeness of classical disjunctive ASP carries over to communicating disjunctive programs. Cautious reasoning is then $\ccoNP$ and $\cco\spolsig{2}$ for communicating normal programs and communicating disjunctive programs, respectively, since this decision problem is the complement of brave reasoning. Finally, the problem of answer set existence is carried over from normal programs and disjunctive programs~\cite{baral:knowledge} and is $\cNP$-hard and $\cco\spolsig{2}$-hard, respectively. Most of these complexity results correspond with classical ASP, with the results from communicating simple programs being notable exceptions;  indeed, for communicating simple programs the communication aspect clearly has an influence on the complexity. Table~\ref{tbl:complexity} summarises the main complexity results.
\begin{table}
  \caption{Completeness results for the main reasoning tasks for a communicating program $\struprog{P}= \set{\srange{Q}{1}{n}}$}
  \label{tbl:complexity}
    \begin{tabular}{r c c c}
    \hline\hline
reasoning task $\rightarrow$ & answer set existence & brave reasoning & cautious reasoning\\
component programs $\downarrow$\\
\hline
simple program & \cP & \cNP & \ccoNP\\
normal program & \cNP & \cNP & \ccoNP\\
disjunctive program & \spolsig{2} & \spolsig{2} & \cco\spolsig{2}\\
\hline\hline
    \end{tabular}
\end{table}

To conclude, given a communicating normal program $\struprog{P}$, there is an easy (linear) translation that transforms $\struprog{P}$ into a normal program $P'$ such that the answer sets of $P'$ corresponds to the answer sets of $\struprog{P}$.

\begin{proposition}
  Let $\struprog{P}$ be a communicating normal program. Let $P'$ be the normal program defined as follows. For every $\ask{Q}{a} \in \hbase{\struprog{P}}$ we add the following rules to $P'$:
  \begin{align}
    \arule{guess(\sask{Q}{a})}{\naf not\_guess(\sask{Q}{a})}&&\arule{not\_guess(\sask{Q}{a})}{\naf guess(\sask{Q}{a})}\nonumber\\
    \arule{}{guess(\sask{Q}{a}), \naf \sask{Q}{a}}&&\arule{}{not\_guess(\sask{Q}{a}), \sask{Q}{a}}\label{sim:guess}.
  \end{align}
 Furthermore, for every normal communicating rule $\struprog{P}$ which is of the form $r = \arule{\ask{Q}{a}}{body}$, we add the rule $\arule{\sask{Q}{a}}{body'}$ to $P'$ with $\sask{Q}{a}$ a literal. We define $body'$, which is a set of extended literals, as follows: 
\begin{align}
  body' &= \condset{\sask{Q}{b}}{\ask{Q}{b} \in body}\nonumber\\
  &\cup \condset{guess(\sask{R}{c})}{\ask{R}{c} \in body, Q \neq R}\nonumber\\
  &\cup \condset{\naf \sask{S}{d}}{(\naf \ask{S}{d}) \in body}.\label{sim:body}
\end{align}
We have that $M=\condset{\ask{Q}{a}}{\sask{Q}{a} \in M'}$ is an answer set of \struprog{P} if and only if $M'$ is an answer set of $P$.
\end{proposition}
\begin{proof}
 (sketch) The essential difference between a normal program and a communicating program is in the reduct. More specifically, the difference is in the treatment of situated literals of the form $\ask{R}{l}$ which are not $Q$-local. Indeed, such literals can, like naf-literals, be guessed and verified whether or not they are stable, \ie verified whether or not the minimal model of the reduct corresponds to the initial guess. It can readily be seen that this behaviour is mimicked by the rules in \eqref{sim:guess}. The first two rules guess whether or not some situated literal $\ask{Q}{a}$ is true, while the last two rules ensure that our guess is stable; \ie we are only allowed to guess $\ask{Q}{a}$ when we are later on actually capable of deriving $\ask{Q}{a}$. The purpose of \eqref{sim:body} is then to ensure that guessing of situated literals is only used when the situated literal in question is not $Q$-local and is not preceded by negation-as-failure.
\end{proof}

%%%%%%%%%%%%%%%%%%%%%%%%%%%%%
%%%%%%%%%%%%%%%%%%%%%%%%%%%%%
%%%
%%%   FOCUSED ANSWER SETS
%%%
%%%%%%%%%%%%%%%%%%%%%%%%%%%%%
%%%%%%%%%%%%%%%%%%%%%%%%%%%%%

\section{Multi-Focused Answer Sets}\label{sec:focused}
Answer set semantics are based on the idea of stable minimal models. When dealing with agents that can communicate, it becomes unclear how we should interpret the notion of minimality. One option is to assume global minimality, \ie we minimise over the conclusions of all the agents in the network. This is the approach that was taken in \thesec{simulating}. Another option is to assume minimality on the level of a single agent. Since it is not always possible to find a model that is minimal for all individual agents, the order in which we minimise over the agents matters, as the next example illustrates. 

\begin{example}
An employee (`$E$') needs a new printer (`$P$'). She has a few choices (loud or silent, stylish or dull), preferring silent and stylish. Her manager~(`$M$') insists that it is a silent printer. Her boss~(`$B$') does not want an expensive printer, \ie one that is both silent and stylish. We~can consider the communicating normal program $\struprog{P} = \set{E, M, B}$ with:
\begin{align}
  \arule{\ask{P}{stylish}&}{\naf \ask{P}{dull}}&\arule{\ask{P}{dull}&}{\naf \ask{P}{stylish}}\label{rules:appearance}\\
  \arule{\ask{P}{silent}&}{\naf \ask{P}{loud}}&\arule{\ask{P}{loud}&}{\naf \ask{P}{silent}}\label{rules:noise}\\
  \arule{\ask{E}{undesired}&}{\ask{P}{dull}}&\arule{\ask{E}{undesired}&}{\ask{P}{loud}}\label{rules:employee}\\
  \arule{\ask{M}{undesired}&}{\ask{P}{loud}}\label{rules:manager}\\
  \arule{\ask{B}{expensive}&}{\ask{P}{stylish}, \ask{P}{silent}}\label{rules:boss}.
\end{align}
The rules in \eqref{rules:appearance} and \eqref{rules:noise} encode the four possible printers and the rules in \eqref{rules:employee}, \eqref{rules:manager} and \eqref{rules:boss} encode the inclinations of the employee, manager and boss, respectively. The answer sets of this program, \ie those with global minimality, are
\begin{align*}
  M_1 &= \set{\ask{P}{stylish}, \ask{P}{silent}, \ask{B}{expensive}}\\
  M_2 &= \set{\ask{P}{stylish}, \ask{P}{loud}, \ask{E}{undesired}, \ask{M}{undesired}}\\
  M_3 &= \set{\ask{P}{dull}, \ask{P}{loud}, \ask{E}{undesired}, \ask{M}{undesired}}\\
  M_4 &= \set{\ask{P}{dull}, \ask{P}{silent}, \ask{E}{undesired}}
\end{align*}
The answer sets that are minimal for agent $B$ are $M_2, M_3$ and $M_4$, \ie the answer sets that do not contain $\ask{B}{expensive}$. The only answer set that is minimal for agent $E$ is $M_1$, \ie the one that does not contain $\ask{E}{undesired}$. Hence when we determine local minimality for communicating ASP, the order in which we consider the agents is important as it induces a priority over them, \ie it makes some agents more important than others. In this example, if the boss comes first, the employee no longer has the choice to pick $M_1$. This leaves her with the choice of either a dull or a loud printer, among which she has no preferences. Since the manager prefers a silent printer, when we first minimise over `$B$' and then minimise over `$M$' and `$E$' (we may as well minimise over `$E$' and then `$M$', as `$E$' and `$M$' have no conflicting preferences) we end up with the unique answer set~$M_4$.
\end{example}

In this section, we formalise such a communication mechanism. We extend the semantics of communicating programs in such a way that it becomes possible to focus on a sequence of component programs. As such, we can indicate that we are only interested in those answer sets that are successively minimal with respect to each respective component program. The underlying intuition is that of leaders and followers, where the decisions that an agent can make are limited by what its leaders have previously decided.

\begin{definition}\label{def:focused}
  Let $\struprog{P}$ be a communicating normal program and $\set{\srange{Q}{1}{n}} \subseteq \struprog{P}$ a set of component programs. A \concept{$\vect{\srange{Q}{1}{n}}$-focused answer set} of \struprog{P} is defined recursively as follows:
\begin{list}{\labelitemi}{\leftmargin=2em} \setlength\itemsep{0.1in}
\item $M$ is a $\vect{\srange{Q}{1}{n}}$-focused answer set of \struprog{P} and there are no\\
${}^{}$ $\vect{\srange{Q}{1}{n-1}}$-focused answer sets $M'$ of \struprog{P} such that $\proj{M'}{Q_n} \subset \proj{M}{Q_n}$;
\item a $\vect{}$-focused answer set of \struprog{P} is any answer set of \struprog{P}.
\end{list}
\end{definition}

In other words, we say that $M$ is a $\vect{\srange{Q}{1}{n}}$-focused answer set of \struprog{P} if and only if $M$ is minimal among all $\vect{\srange{Q}{1}{n-1}}$-focused answer sets \wrt the projection on $Q_n$.

\begin{example}\label{ex:subsetminimal}
Consider the communicating normal program $\struprog{P} = \set{Q,R,S}$ with the rules 
\begin{align*}
\arule{\ask{Q}{a}&}{}&\arule{\ask{R}{b}&}{\ask{S}{c}}&\arule{\ask{S}{a}&}{}\\
\arule{\ask{Q}{b}&}{\naf \ask{S}{d}}&\arule{\ask{R}{a}&}{\ask{S}{c}}&\arule{\ask{S}{c}&}{\naf \ask{S}{d}, \naf \ask{R}{c}}\\
\arule{\ask{Q}{c}&}{\ask{R}{c}}&\arule{\ask{R}{a}&}{\ask{S}{d}}&\arule{\ask{S}{c}&}{\naf \ask{S}{c}, \naf \ask{R}{c}}\\
&&\arule{\ask{R}{c}&}{\naf \ask{R}{a}}
\end{align*}
The communicating normal program $\struprog{P}$ has three answer sets, namely 
\begin{align*}
M_1 &= \ask{Q}{\set{a,b,c}} \cup \ask{R}{\set{c}} \cup \ask{S}{\set{a}}\\
M_2 &= \ask{Q}{\set{a,b}} \cup \ask{R}{\set{a,b}} \cup \ask{S}{\set{a,c}}\\
M_3 &= \ask{Q}{\set{a}} \cup \ask{R}{\set{a}} \cup \ask{S}{\set{a,d}}.
\end{align*}

The only $\vect{R,S}$-focused answer set of $\struprog{P}_{\ref{ex:subsetminimal}}$ is $M_1$. Indeed, since $\set{a} = \proj{(M_3)}{R} \subset \proj{(M_2)}{R} = \set{a,b}$ we find that $M_2$ is not a $\vect{R}$-focused answer set. Furthermore $\set{a} = \proj{(M_1)}{S} \subset \proj{(M_3)}{S} = \set{a,d}$, hence $M_3$ is not an $\vect{R, S}$-focused answer set.
\end{example}

\begin{proposition}\label{prop:one-focused-is-polynomial}
Let \struprog{P} be a communicating simple program. We then have:
\begin{list}{\labelitemi}{\leftmargin=2em}
  \item there always exists at least one $\vect{Q_1, ..., Q_n}$-focused answer set of $\struprog{P}$;
  \item we can compute this $\vect{Q_1, ..., Q_n}$-focused answer set in polynomial time. 
\end{list}
\end{proposition}

To investigate the computational complexity of multi-focused answer sets we now show how the validity of quantified boolean formulas (QBF) can be checked using multi-focused answer sets of communicating ASP programs.

\begin{definition}\label{def:simulation-QBF}
  Let $\phi = \exists X_1 \forall X_2 ... \mathrm{\Theta} X_n \cdot p(X_1, X_2, \cdots X_n)$ be a QBF where $\mathrm{\Theta} = \forall$ if $n$ is even and $\mathrm{\Theta} = \exists$ otherwise, and $p(X_1, X_2, \cdots X_n)$ is a formula of the form $\srangec{\theta}{1}{m}{\vee}$ in disjunctive normal form over $\srangec{X}{1}{n}{\cup}$ with $X_i$, $1 \leq i \leq n$, sets of variables and where each $\theta_t$ is a conjunction of propositional literals. We define $Q_0$ as follows:
\begin{align}
  Q_0 = &\condset{\arule{x}{\naf \neg x}, \arule{\neg x}{\naf x}}{x \in \srangec{X}{1}{n}{\cup}}\label{eq:assignments}\\
  & \cup \condset{\arule{sat}{\ask{Q_0}{\theta_t}}}{1 \leq t \leq m}\label{eq:satisfied}\\
  &\cup \set{\arule{\neg sat}{\naf sat}}.
\end{align}
For $1 \leq j \leq n-1$ we define $Q_j$ as follows:
\begin{align}
  Q_j = &\condset{\arule{x}{\ask{Q_0}{x}}, \arule{\neg x}{\ask{Q_0}{\neg x}}}{x \in (\srangec{X}{1}{n-j}{\cup})}\label{eq:assume}\\
  & \cup \begin{cases}\label{eq:saturate}
  \set{\arule{\neg sat}{\ask{Q_0}{\neg sat}}}&\text{if $(n-j)$ is even}\\
  \set{\arule{sat}{\ask{Q_0}{sat}}}&\text{if $(n-j)$ is odd}.
  \end{cases}
\end{align}
The communicating normal program corresponding with $\phi$ is $\struprog{P}_{\phi} = \set{\srange{Q}{0}{n-1}}$.

 For a QBF of the form $\phi = \forall X_1 \exists X_2 ... \mathrm{\Theta} X_n \cdot p(X_1, X_2, \cdots X_n)$ where $\mathrm{\Theta} = \exists$ if $n$ is even and $\mathrm{\Theta} = \forall$ otherwise and $p(X_1, X_2, \cdots X_n)$ once again a formula in disjunctive normal form, the simulation only changes slightly. Indeed, only the conditions in~\eqref{eq:saturate} are swapped. 
 \end{definition}

\begin{example}\label{ex:simulation-QBF}
  Given the QBF $\phi = \exists x \forall y \exists z \cdot (x \wedge y) \vee (\neg x \wedge y \wedge z) \vee (\neg x \wedge \neg y \wedge \neg z)$, the communicating normal program $\struprog{P}$ corresponding with the QBF $\phi$ is defined as follows:
  \begin{align*}
    \arule{\ask{Q_0}{x}&}{\naf \neg x}&\arule{\ask{Q_0}{y}&}{\naf \neg y}&\arule{\ask{Q_0}{z}&}{\naf \neg z}\\
    \arule{\ask{Q_0}{\neg x}&}{\naf x}&\arule{\ask{Q_0}{\neg y}&}{\naf y}&\arule{\ask{Q_0}{\neg z}&}{\naf z}\\
    \arule{\ask{Q_0}{sat}&}{x,y}&\arule{\ask{Q_0}{sat}&}{\neg x, y, z}&\arule{\ask{Q_0}{sat}&}{\neg x, \neg y, \neg z}\\
    \arule{\ask{Q_0}{\neg sat}&}{\naf sat}\\~\\
    \arule{\ask{Q_1}{x}&}{\ask{Q_0}{x}}&\arule{\ask{Q_1}{y}&}{\ask{Q_0}{y}}\\
    \arule{\ask{Q_1}{\neg x}&}{\ask{Q_0}{\neg x}}&\arule{\ask{Q_1}{\neg y}&}{\ask{Q_0}{\neg y}}&\arule{\ask{Q_1}{\neg sat}&}{\ask{Q_0}{\neg sat}}\\~\\
    \arule{\ask{Q_2}{x}&}{\ask{Q_0}{x}}&\arule{\ask{Q_2}{\neg x}&}{\ask{Q_0}{\neg x}}&\arule{\ask{Q_2}{sat}&}{\ask{Q_0}{sat}}
  \end{align*}
\end{example}

The communicating normal program in \themyex{simulation-QBF} can be used to determine whether the QBF $\phi$ is valid. First, note that the rules in~\eqref{eq:assignments} generate all possible truth assignments of the variables, \ie all possible propositional interpretations. The rules in \eqref{eq:satisfied} ensure that `$sat$' is true exactly for those interpretations that satisfy the formula $p(X_1, \srange{X}{2}{n})$. 

Intuitively, the component programs $\set{\srange{Q}{1}{n-1}}$ successively bind fewer and fewer variables. In particular, focusing on $\srange{Q}{1}{n-1}$ allows us to consider the binding of the variables in $\srange{X}{n-1}{1}$, respectively. Depending on the rules from~\eqref{eq:saturate}, focusing on $Q_i$ allows us to verify that either some or all of the assignments of the variables in $X_{n-j}$ make the formula $p(\srange{X}{1}{n})$ satisfied, given the bindings that have already been determined by the preceding components. We now prove that the QBF $\phi$ is satisfiable iff $\ask{Q_0}{sat}$ is true in some $\vect{\srange{Q}{1}{n-1}}$-focused answer set of the corresponding program.

\begin{proposition}\label{prop:qbf}
  Let $\phi$ and \struprog{P} be as in \themydef{simulation-QBF}.
  We have that a QBF $\phi$ of the form $\phi = \exists X_1 \forall X_2 ... \mathrm{\Theta} X_n \cdot p(X_1, X_2, \cdots X_n)$ is satisfiable if and only if $\ask{Q_0}{sat}$ is true in some $\vect{\srange{Q}{1}{n-1}}$-focused answer set of $\struprog{P}$.  Furthermore, we have that a QBF $\phi$ of the form $\phi = \forall X_1 \exists X_2 ... \mathrm{\Theta} X_n \cdot p(X_1, X_2, \cdots X_n)$ is satisfiable if and only if $\ask{Q_0}{sat}$ is true in all $\vect{\srange{Q}{1}{n-1}}$-focused answer sets of $\struprog{P}$. 
\end{proposition}

\begin{corollary}\label{cor:sigma}
  Let \struprog{P} be a communicating normal program with $Q_i \in \struprog{P}$. The problem of deciding whether there exists a $\vect{\srange{Q}{1}{n}}$-focused answer set $M$ of $\struprog{P}$ such that  $\ask{Q_i}{l} \in M$ (brave reasoning) is \spolsig{n+1}-hard.
\end{corollary}

\begin{corollary}\label{cor:pi}
  Let \struprog{P} be a communicating normal program with $Q_i \in \struprog{P}$. The problem of deciding whether all $\vect{\srange{Q}{1}{n}}$-focused answer sets contain $\ask{Q_i}{l}$ (cautious reasoning) is \spolpi{n+1}-hard.
\end{corollary}

In addition to these hardness results, we can also establish the corresponding membership results.

\begin{proposition}\label{prop:focused-complete}
  Let \struprog{P} be a communicating normal program with $Q_i \in \struprog{P}$. The problem of deciding whether there exists a $\vect{\srange{Q}{1}{n}}$-focused answer set $M$ of $\struprog{P}$ such that  $\ask{Q_i}{l} \in M$ (brave reasoning) is in \spolsig{n+1}.
\end{proposition}

Since cautious reasoning is the complementary problem of brave reasoning it readily follows that cautious reasoning is in $\cco\spolsig{n=1}$. Now that we have both hardness and membership results, we readily obtain the following corollary.

\begin{corollary}
 Let \struprog{P} be a communicating normal program with $Q_i \in \struprog{P}$. The problem of deciding whether $\ask{Q_i}{l} \in M$ with $M$ a $\vect{\srange{Q}{1}{n}}$-focused answer set of $\struprog{P}$ is \spolsig{n+1}-complete.
\end{corollary}

The next corollary shows that the complexity remains the same when going from normal component programs to simple component programs.
% It is important to note that this corollary is based on complexity results, not directly on the result that a QBF can be simulated. If we were to simulate a QBF with component simple programs, we would need to use another simulation than the one proposed in \themydef{qbf}. Indeed, we need to build in the requirement of the guess to be total in one of the component simple programs. This can easily be done by introducing new rules such that we have for some $Q_i$ that $\ask{Q_i}{total}$ whenever our guess is total. Furthermore, we would need to update the rules that indicate whether the QBF is satisfied or not to only be true when we have a guess that is total.

\begin{proposition}\label{prop:simple-complete}
 Let \struprog{P} be a communicating simple program with $Q_i \in \struprog{P}$. The problem of deciding whether there exists a $\vect{\srange{Q}{1}{n}}$-focused answer set $M$ of $\struprog{P}$ such that $\ask{Q_i}{l} \in M$ (brave reasoning) is in \spolsig{n+1}.
\end{proposition}

Finally, we also have a result for communicating disjunctive programs instead of communicating normal programs. 

\begin{proposition}\label{prop:disjunctive-complete}
 Let \struprog{P} be a communicating disjunctive program with $Q_i \in \struprog{P}$. The problem of deciding whether $\ask{Q_i}{l} \in M$ with $M$ a $\vect{\srange{Q}{1}{n}}$-focused answer set of $\struprog{P}$ is in \spolsig{n+2}.
\end{proposition}

\noindent Table~\ref{tbl:results} summarises the membership results for brave reasoning that were discussed in this section. 
\begin{table}
  \caption{Membership results for brave reasoning with (multi-focused) answer sets of the communicating program $\struprog{P}= \set{\srange{Q}{1}{n}}$}
  \label{tbl:results}
    \begin{tabular}{r c c c}
    \hline\hline
form of communication $\rightarrow$ & none & situated literals & multi-focused\\
type of component program $\downarrow$\\
\hline
simple program & \cP & \cNP & \spolsig{n+1}\\
normal program & \cNP & \cNP & \spolsig{n+1}\\
disjunctive program & \spolsig{2} & \spolsig{2} & \spolsig{n+2}\\
\hline\hline
    \end{tabular}
\end{table}

Due to the extra expressiveness and complexity of multi-focused answer sets, it is clear that no translation to classical ASP is possible. Possible future implementations may, however, be based on a translation to other PSPACE complete problems such as QBF formulas or modal logics. A translation to QBF formulas seems to be the most natural, especially since the proof of the complexity of multi-focused answer sets involves reducing QBF formulas to multi-focused answer sets. However, any such translation falls beyond the scope is this paper and is the subject of future research.

\section{Case Study: subset-minimal abductive diagnosis}\label{sec:casestudy}
In this section we work out an example that highlights the usefulness of multi-focused answer sets. The use of multi-focused answer sets has already proven itself useful in modeling problems where one can use a negotiation paradigm, \eg in~\cite{bauters:modeling}. However, the main goal in~\cite{bauters:modeling} was to show that such a paradigm is possible, rather than actually trying to encode a problem that is known to be more complex that $\spolsig{2}$. Though a lot of interesting problems are indeed in \cP, \cNP\ or \spolsig{2}, there are still some important problems that are even higher up in the polynomial hierarchy. One such a problem is a special form of abductive diagonistics. An abductive diagnostic problem is encoded as a triple $\langle H, T, O \rangle~$\cite{eiter:diagnosis}, where $H$ is a set of atoms referred to as hypotheses, $T$ is an ASP program referred to as the theory and $O$ is a set of literals referred to as observations. Intuitively, the theory $T$ describes the dynamics of the system, the observations $O$ describe the observed state of the system and the hypotheses $H$ try to explain these observations within the theory. The goal in subset-minimal abductive diagnosis is to find the minimal set of hypotheses that explain the observation. That is, we want to find the minimal set of hypotheses such that $O \subseteq M$ with $M$ an answer set of $T \cup H$. Subset-minimal abductive diagnostics over a theory consisting of a disjunctive program is a problem in \spolsig{3} and hence we cannot rely on classical ASP to find the solutions to this problem. However, as we will see in the next example, we can easily solve this problem using multi-focused answer sets.

\begin{example}[Adapted from \cite{eiter:diagnosis}]\label{ex:diagnosis}
Consider an electronic circuit, as in \thefig{diagnosis}, where we have a power source, a control lamp, three hot-plates wired in parallel and a fuse to protect each hot-plate. It is known that some of the fuses are sensitive to high current and may consequently blow, but it is not known which fuses. Furthermore, plate A sits near a source of water (\eg a tap). If water comes into contact with plate A, this causes a short circuit which blows the nearest fuse, \ie fuse A, to prevent any damage.

\begin{figure}[ht]
  \includegraphics[scale=0.5]{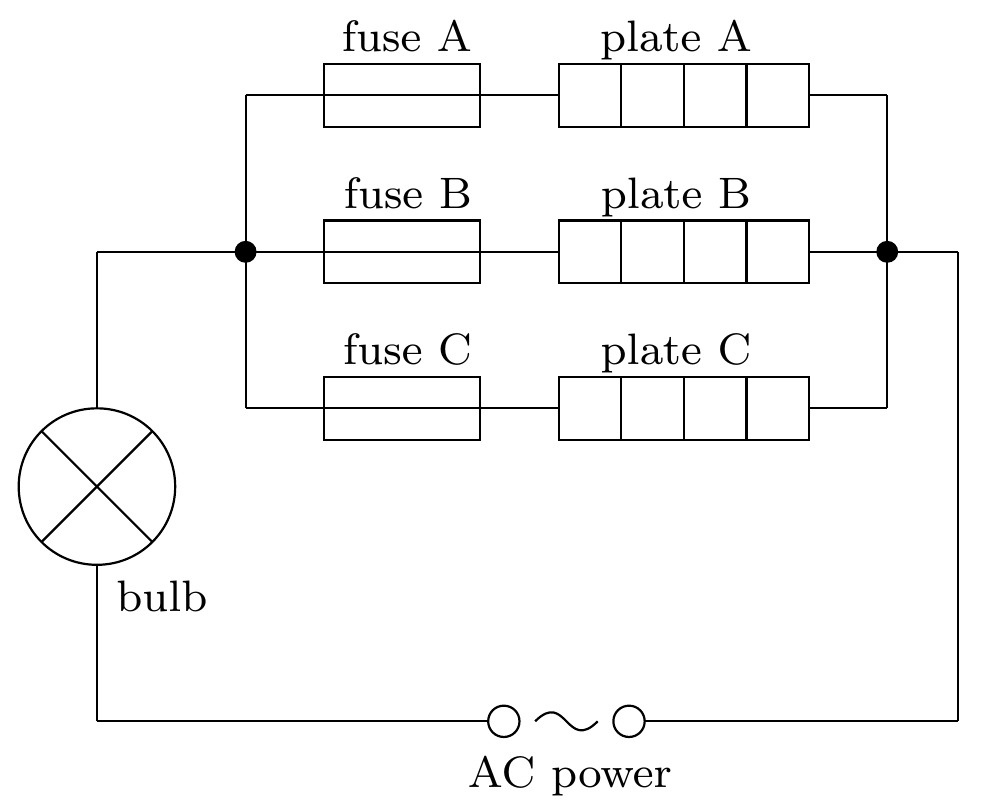}
  \caption{Schematics of the electronic circuit we want to diagnose.}
  \label{fig:diagnosis}
\end{figure}
\end{example}

Upon inspection, we find that the control lamp is on and that plate A feels cold to the touch. We want to find the subset minimal diagnoses that would explain the problem, \ie we want to find the minimal causes that can explain this situation. 

First we need to describe the theory, \ie the schematics. The theory describes the dynamics of the system and thus also how the system may fail. We can describe the theory as follows. For starters, a melted fuse can be caused by a high current, or, for fuse A, due to a hazardous water leak:
{\allowdisplaybreaks
\begin{align*}
  \sarule{\ask{Q}{melted\us A} ; \ask{Q}{melted\us B} ; \ask{Q}{melted\us C}}{\ask{Q}{high}}\\
  \sarule{\ask{Q}{melted\us A}}{\ask{Q}{leak}}.
  \intertext{Furthermore, under a number of conditions the control light will be off:}
  \sarule{\ask{Q}{light\us off}}{\ask{Q}{power\us off}}\\
  \sarule{\ask{Q}{light\us off}}{\ask{Q}{broken\us bulb}}\\
  \sarule{\ask{Q}{light\us off}}{\ask{Q}{melted\us A}, \ask{Q}{melted\us B}, \ask{Q}{melted\us C}}.
  \intertext{Then we describe under what conditions each plate will be hot:}
  \sarule{\ask{Q}{hot\us plateA}}{\naf \ask{Q}{melted\us A}, \naf \ask{Q}{power\us off}}\\
  \sarule{\ask{Q}{hot\us plateB}}{\naf \ask{Q}{melted\us B}, \naf \ask{Q}{power\us off}}\\
  \sarule{\ask{Q}{hot\us plateC}}{\naf \ask{Q}{melted\us C}, \naf \ask{Q}{power\us off}}.
\end{align*}}
We now encode the hypotheses. We have a number of causes, each of which may by itself or in conjunction with other causes explain our observation. In total, we have four causes. The power can be off ($\mathit{power\us off}$), the light bulb might be broken ($\mathit{broken\us bulb}$), there may have been a high current ($\mathit{high}$) and/or a water leak may have occurred ($\mathit{leak}$). We describe all these hypotheses as follows:
{\allowdisplaybreaks
\begin{align*}
  \sarule{\ask{Q}{power\us off}}{\naf \ask{Q}{no\us power\us off}}\\
  \sarule{\ask{Q}{no\us power\us off}}{\naf \ask{Q}{power\us off}}\\
  \sarule{\ask{Q}{broken\us bulb}}{\naf \ask{Q}{no\us broken\us bulb}}\\
  \sarule{\ask{Q}{no\us broken\us bulb}}{\naf \ask{Q}{broken\us bulb}}\\
  \sarule{\ask{Q}{high}}{\naf \ask{Q}{no\us high}}\\
  \sarule{\ask{Q}{no\us high}}{\naf \ask{Q}{high}}\\
  \sarule{\ask{Q}{leak}}{\naf \ask{Q}{no\us leak}}\\
  \sarule{\ask{Q}{no\us leak}}{\naf \ask{Q}{leak}}.
\end{align*}}
It is easy to see that these rules in $Q$ encode all possible subsets of hypotheses that may have occurred. We then add rules to a separate component program $H$ which merely relays the information on the set of hypotheses that we chose. The reason for this separate component program $H$ is that we can now minimise over the set of hypotheses that is assumed, simply by focusing on $H$.
\begin{align*}
  \sarule{\ask{H}{power\us off}}{\ask{Q}{power\us off}}\\
  \sarule{\ask{H}{broken\us bulb}}{\ask{Q}{broken\us bulb}}\\
  \sarule{\ask{H}{high\us current}}{\ask{Q}{high}}\\
  \sarule{\ask{H}{water\us leak}}{\ask{Q}{leak}}
\end{align*}
Finally, we model the observation. We observe that the control light is on and that plate A is cold. In other words, we obtain the rules (which encode constraints):
\begin{align*}
  \sarule{\ask{Q}{contradiction}}{\naf \ask{Q}{contradiction}, \ask{Q}{light\us off}}\\
  \sarule{\ask{Q}{contradiction}}{\naf \ask{Q}{contradiction}, \ask{Q}{hot\us plateA}}
\end{align*}
which intuitively tell us that we cannot have that the light is off, nor can we have that plate A is hot. 

The $\vect{H}$-focused answer sets give us the subset minimal abductive diagnoses. It is easy to see that the focus on $H$ is needed to minimise over the hypotheses. The program $\struprog{P} = \set{Q,H}$ has two $\vect{H}$-focused answer sets $M_1$ and $M_2$, both containing $M_{\mathrm{shared}} = \set{\ask{Q}{\mathit{no\us power\us off}}, \ask{Q}{\mathit{no\us broken\us bulb}}, \ask{Q}{\mathit{hot\us plateB}}, \ask{Q}{\mathit{hot\us plateC}}}$:
\begin{align*}
  M_1 = M_{\mathrm{shared}} &\cup \set{\ask{Q}{melted\us A}, \ask{Q}{no\us leak}, \ask{Q}{high}, \ask{H}{high}}\\
  M_2 = M_{\mathrm{shared}} &\cup \set{\ask{Q}{melted\us A}, \ask{Q}{leak}, \ask{Q}{no\us high}, \ask{H}{leak}}. 
\end{align*}
Hence the minimal sets of hypotheses that support our observation, \ie $\proj{M}{H}$ with $M$ an $\vect{H}$-focused answer set, are that either there was a high current (which melted fuse A) or there was a water leak (which also melted fuse A).

\section{Related Work}\label{sec:related}
A large body of research has been devoted to combining logic programming with multi-agent or multi-context ideas, with various underlying reasons. One reason for such a combination is that the logic can be used to describe the (rational) behaviour of the agents in a multi-agent network, as in~\cite{acqua:communicating}. Alternately, it can be used to combine different flavours of logic programming languages~\cite{luo:cooperation,eiter:combining}. It can also be used to externally solve tasks for which ASP is not suited, while remaining in a declarative framework~\cite{eiter:dlvhex}. As a final example, it can be used as a form of cooperation, where multiple agents or contexts collaborate to solve a difficult problem~\cite{devos:laima,nieuwenborgh:hierarchical}. The approach in this paper falls in the last category and is concerned with how the collaboration of different ASP programs affects the expressiveness of the overall system. 

Important work has been done in the domain of multi-context systems (MCS) and multi-agent ASP to enable collaboration between the different contexts/ASP programs. We discuss some of the more prominent work in these areas.

The work of~\cite{roelofsen:minimal} proposes an extension of multi-context systems or MCSs~\cite{giunchiglia:multilanguage} that allows MCSs to reason about absent information, \ie they introduce non-monotonicity in the context of MCSs. The idea of a MCS is that we have a number of contexts that each have access to only a subset of the available information. Each context has a local model and reasoning capabilities, but there is also an information flow defined by the system between the different contexts. It is this idea that was later adopted in the ASP community and in our paper in particular. 

Our paper has a comparable syntax as~\cite{roelofsen:minimal} but rather different semantics. The semantics in~\cite{roelofsen:minimal} are closely related to the well-founded semantics~\cite{gelder:well}, while our semantics are closer in spirit to the stable model semantics~\cite{gelfond:stablemodel}. Another point where our semantics differ is that we allow a restricted circular explanation of why a literal is true, if that circular explanation is due to our reliance on other component programs. This particular form of circular reasoning has been identified in~\cite{buccafurri:logic} as a requirement in the representation of social reasoning. 

The work in~\cite{brewka:contextual} further extends the work in~\cite{roelofsen:minimal} and addresses a number of problems and deficiencies. The paper is, to the best of our knowledge, the first to offer a syntactical rather than semantical description of communication in multi-context systems, making it easier to implement an actual algorithm. A number of interesting applications of contextual frameworks, including information fusion, game theory and social choice theory are highlighted in the paper. Lastly, the paper identifies that the complexity of the main reasoning task is on the second level of the polynomial hierarchy. 

Along similar lines~\cite{brewka:equilibria} combines the non-monotonicity from~\cite{roelofsen:minimal} with the heterogeneous approach presented in~\cite{giunchiglia:multilanguage} into a single framework for heterogenous non-monotonic multi-context reasoning. The work in~\cite{brewka:equilibria} introduces several notions of equilibria, including minimal and grounded equilibria. In our approach, local reasoning is captured by grounded equilibria (which does not allow circular explanations) while communicating with other component programs is captured by the weaker minimal equilibria. The work in~\cite{brewka:equilibria} offers various membership results on deciding the existence of an equilibrium and is one of the first to note that multi-context systems, due to the nature of the bridge rules/situated literals, can be non-monotonic even if all the logics in the component programs themselves are monotonic. 

Work on a distributed solver for heterogenous multi-context systems commenced in~\cite{daotran:distributed}. While solvers exist to compute multi-context systems locally, this is the first work to consider an algorithm which is both distributed (\ie no shared memory) and modular (\ie computation starting from partial models). When the context under consideration uses \eg ASP, loop formulas can be devised which allow bridge rules to be compiled into local classical theories. It is then possible to use SAT solvers to compute the grounded equilibria of the heterogenous multi-context system. Later work in~\cite{drescher:symmetry} improved on the idea by offering a mechanism to identify and break symmetries (\ie permutations of belief which result in identical knowledge). As such, the solver need never visit two points in the search space that are symmetric, thus potentially offering a considerable speedup. Experimental results show that the solution space can indeed be (significantly) compressed. A similar idea might be used to compute answer sets of a communicating ASP program in a distributed fashion. Indeed, such answer sets are closely related to the idea of minimal equilibria from~\cite{brewka:equilibria}. A few modifications should nonetheless be made. For example, the Herbrand base needs to be redefined in a way that is safe in such a distributed setting, \eg by only taking situated literals into account that occur in a given component program. Optimizations to the distributed algorithm also seem likely to be applicable to the setting of communicating ASP. On the other hand, it does not seem to be straightforward to extend these ideas to compute multi-focused answer sets in a distributed fashion. 

One of the most recent extensions to multi-context systems are managed multi-context systems or mMCS~\cite{brewka:managed}. Normally, bridge rules can only be used to pass along information which allows for \eg selection and abstraction of information between contexts. In an mMCS, however, additional operations on knowledge bases can be freely defined. For example, operations may be defined that remove or revise information. Such operations are performed by the context itself, \ie by the legacy system that is used such as ASP, but mMCS allow to cope with this additional functionality in a principled way. As one would expect, adding such complex operations increases the expressiveness of the resulting system considerably. Our work, on the other hand, only allows for information to be passed along. By varying the way that the communication works, we achieved a comparable expressiveness.

We now direct our attention to work done within the ASP community. The ideas presented in this paper are related to HEX programs~\cite{eiter:uniform} in which ASP is extended with higher-order predicates and external atoms. These external atoms allow to exchange knowledge in a declarative way with external sources that may implement functionality which is inconvenient or impossible to encode using current answer set programming paradigms. Application-wise, HEX is mainly proposed as a tool for non-monotonic semantic web reasoning under the answer set semantics. Hence HEX is not primarily targeted at increasing the expressiveness, but foremost at extending the applicability and ease of use of ASP.

Two other important works in the area of multi-agent ASP are~\cite{devos:laima} and~\cite{nieuwenborgh:hierarchical}. In both~\cite{devos:laima} and~\cite{nieuwenborgh:hierarchical} a multi-agent system is developed in which multiple agents/component programs can communicate with each other. Most importantly from the point of view of our work, both approaches use ASP and have agents that are quite expressive in their own right. Indeed, in ~\cite{devos:laima} each agent is an Ordered Choice Logic Program (OCLP)~\cite{brain:oclp} and in~\cite{nieuwenborgh:hierarchical} each agent uses the extended answer set semantics.

The framework introduced in~\cite{devos:laima} is called LAIMA. Each of the agents is an OCLP. The agents can communicate with whoever they want and circular communication is allowed (where agent $A$ tells something to agent $B$ which tells something to $A$ \ldots). However, only positive information can be shared and the authors do not look at the actual expressiveness of the framework. In~\cite{nieuwenborgh:hierarchical} each agent uses the extended answer set semantics. The network is a linear ``hierarchical'' network (\ie information only flows in one direction), yet they employ the idea of a failure feedback mechanism. Intuitively, a failure feedback mechanism allows the previous agent in a network to revise his conclusion when the conclusion leads to an unresolvable inconsistency for the next agent in the network. It is this mechanism that gives rise to a higher expressiveness, namely \spolsig{n} for a hierarchical network of $n$ agents. Our work is different in that we start from simple  and normal ASP programs for the agents. Our communication mechanism is also quite simple and does not rely on any kind of feedback. Regardless, we obtain a comparable expressiveness. 

We also mention~\cite{dao-tran:modular} where recursive modular non-monotonic logic programs (MLP) under the ASP semantics are considered. The main difference between MLP and our work is that our communication mechanism is parameter-less, \ie the truth of a situated literal is not dependent on parameters passed by the situated literal to the target component program. Our approach is clearly different and we cannot readily mimic the behaviour of the networks presented in~\cite{dao-tran:modular}. Our expressiveness results therefore do not directly apply to MLPs.

Finally, there is an interesting resemblance between multi-focused answer sets and the work on multi-level integer programming~\cite{jeroslow:polynomial}. In multi-level integer programming, different agents control different variables that are outside of the control of the other agents, yet are linked by means of linear inequalities (constraints). The agents have to fix the values of the variables they can control in a predefined order, such that their own linear objective function is optimized. Similarly, in communicating ASP, literals belong to different component programs (agents), and their values are linked through constraints, which in this case take the form of rules.  Again the agents act in a predefined order, but now they try to minimise the set of literals they have to accept as being true, rather than a linear objective function. Though there is an intuitive link, further research is required to make this link between multi-focused answer sets and the work on multi-level integer programming explicit.

%%%%%%%%%%%%%%%%%%%%%%%%%%%%%
%%%%%%%%%%%%%%%%%%%%%%%%%%%%%
%%%
%%%   CONCLUSIONS
%%%
%%%%%%%%%%%%%%%%%%%%%%%%%%%%%
%%%%%%%%%%%%%%%%%%%%%%%%%%%%%

\section{Conclusions}\label{sec:conclusions}
We have systematically studied the effect of adding communication to ASP in terms of expressiveness and computational complexity. We start from simple programs, \ie definite programs extended with true negation. Determining whether a literal belongs to an answer set of a simple program is a problem in \cP. A network of these simple programs, which we call communicating simple programs, is however expressive enough to simulate normal programs. In other words, determining whether a literal belongs to an answer set of a communicating simple program is \cNP-hard. Furthermore, communicating simple programs can also simulate communicating normal programs provided that the resulting answer sets are partially complete, thus showing that adding negation-as-failure to communicating simple programs does not further increase the expressiveness.

We have introduced multi-focused answer sets for communicating programs. The underlying intuition is that of leaders and followers, where the choices available to the followers are limited by what the leaders have previously decided. On a technical level, the problem translates to establishing local minimality for some of the component programs in the communicating program. Since in general it is not possible to ensure local minimality for all component programs, an order must be defined among component programs on which to focus. The result is an increase in expressiveness, where the problem of deciding whether $\ask{Q_i}{l} \in M$ with $M$ a $\vect{\srange{Q}{1}{n}}$-focused answer set of $\struprog{P}$ is \spolsig{n+1}-complete. In our work we thus find that the choice of the communication mechanism is paramount \wrt the expressiveness of the overall system, in addition to the expressiveness of the individual agents. Table~\ref{tbl:results} highlights the membership results for brave reasoning obtained in \thesec{focused}. 

\appendix

\section{Result proofs}
\begin{reproposition}{one-simple-is-polynomial}
Let \struprog{P} be a communicating simple program. We then have that:
\begin{list}{\labelitemi}{\leftmargin=2em}
  \item there always exists at least one answer set of $\struprog{P}$;
  \item there is always a unique answer set of $\struprog{P}$ that is globally minimal;
  \item we can compute this unique globally minimal answer set in polynomial time. 
\end{list}
\end{reproposition}

\begin{proof}
We can easily generalise the immediate consequence operator for (classical) simple programs to the case of communicating simple programs. Specifically, the operator $T_{\struprog{P}}$ is defined \wrt an interpretation $I$ of $\struprog{P}$ as $$T_{\struprog{P}}(I) = I \cup \condset{\ask{Q}{l}}{(\arule{\ask{Q}{l}}{\alpha}) \in Q, Q \in \struprog{P}, \alpha \subseteq I}$$ where $\alpha$ is a set of $\struprog{P}$-situated literals. It is easy to see that this operator is monotone. Together with a result from \cite{tarski:lattice} we know that this operator has a least fixpoint. We use $\struprog{P}^{\star}$ to denote this fixpoint obtained by repeatedly applying $T_{\struprog{P}}$ starting from the empty interpretation. Clearly, this fixpoint can be computed in polynomial time.

We need to verify that $\struprog{P}^{\star}$ is indeed an answer set. Since \struprog{P} is a communicating simple program, we know that the reduct ${Q}^{\struprog{P}^{\star}}$ will only remove rules that contain situated literals $\ask{R}{l}$ that are not $Q$-local with $\ask{R}{l} \notin \struprog{P}^{\star}$. In other words, rules that are not applicable ($\alpha \not\subseteq\struprog{P}^{\star}$) and that contain non-$Q$-local situated literals are removed. Furthermore, remaining situated literals of the form $\ask{R}{l}$ that are not $Q$-local (\ie those where $\ask{R}{l} \in \struprog{P}^{\star}$) are removed from the body of the remaining rules. Hence the remaining rules are all $Q$-local. Notice that the operator $T_{\struprog{P}}$ is clearly an extension of the operator $T_Q$. Indeed, for a component simple program $Q'$ that is $Q'$-local it is easy to verify that if ${(Q')}^{\star} = M'$ then $\proj{({(\struprog{P'})}^{\star})}{Q'} = M'$ with $\struprog{P'} = \set{Q'}$. It then readily follows, since all rules are $Q$-local and therefore independent of all other component programs, that ${\left(Q^{\struprog{P}^{\star}}\right)}^{\star} = \proj{(\struprog{P}^{\star})}{Q}$ for all $Q \in \struprog{P}$. 

So far we found that an answer set exists and that it can be computed in polynomial time. All that remains to be shown is that this answer set is globally minimal. This trivially follows from the way we defined the operator $T_{\struprog{P}}$ since it only makes true the information that is absolutely necessary, \ie the information that follows directly from the facts in the communicating simple program. Hence this is the minimal amount of information that needs to be derived for a set of situated literals to be a model of the communicating simple program at hand and thus the fixpoint $\struprog{P}^{\star}$ is the globally minimal answer set.
\end{proof}

\begin{relemma}{analogousanswerset}
Let $\struprog{P} = \set{\srange{Q}{1}{n}}$ and let $\struprog{P}' = \set{\srange{Q'}{1}{n}, \srange{N}{1}{n}}$ with $\struprog{P}$ a communicating normal program and $\struprog{P}'$ the communicating simple program that simulates \struprog{P} as defined in \themydef{simulateextended}. Let $M$ be an answer set of \struprog{P} and let the interpretation $M'$ be defined~as:
  \begin{align}\label{analogousanswerset}
  \begin{split}
    M' = &\condset{\ask{Q'_i}{a}}{a \in \proj{M}{Q_i}, Q_i \in \struprog{P}}\\
    &\cup~\condset{\ask{Q'_i}{\neg\fresh{ b}}}{b \notin \proj{M}{Q_i}, Q_i \in \struprog{P}}\\
    &\cup~\condset{\ask{N_i}{\neg\fresh{ b}}}{b \notin \proj{M}{Q_i}, Q_i \in \struprog{P}}\\
    &\cup~\condset{\ask{N_i}{\fresh{a}}}{a \in \proj{M}{Q_i}, Q_i \in \struprog{P}}.
    \end{split}
  \end{align}
  For each $i \in \set{\range{1}{n}}$ it holds that $(Q'_i+)^{M'} = \condset{\arule{l}{\alpha'}}{\arule{l}{\alpha} \in Q^M_i}$ with $Q'_i+$ the set of rules defined in \eqref{simulation-Q-pos} with $\alpha' = \condset{\ask{Q'_{i}}{b}}{\ask{Q_{i}}{b} \in \alpha}$.
\end{relemma}

\begin{proof}
To prove this, we first show that any rule of the form $(\arule{l}{\alpha}) \in Q^M_i$ reappears in $(Q'_i+)^{M'}$ under the form $(\arule{l}{\alpha'})$ for any $i \in \set{\range{1}{n}}$. The second step, showing that the converse also holds, can then be done in an analogous way.

Suppose $(\arule{l}{\alpha}) \in Q^M_i$ for some $i \in \set{\range{1}{n}}$. By the definition of the reduct we know that there is some rule of the form $(\arule{l}{\alpha \cup \naf \beta \cup \gamma}) \in Q_i$ such that $\beta \cap M = \emptyset$ and $\gamma \subseteq M$ is a set of situated literals of the form $\ask{Q_j}{d}$ with $i \neq j$, $1 \leq j \leq n$. From \themydef{simulateextended}, we know that the communicating normal rule $(\arule{\ask{Q_i}{l}}{\alpha \cup \naf \beta \cup \gamma})$ is transformed into the rule $(\arule{\ask{Q'_i}{l}}{\alpha' \cup \beta' \cup \gamma'})$ with $\alpha' = \condset{\ask{Q'_i}{b}}{\ask{Q_i}{b} \in \alpha}$, $\beta' = \condset{\ask{N_k}{\neg\fresh{ c}}}{\ask{Q_k}{c} \in \beta, k \in \set{\range{1}{n}}}$ and $\gamma' = \condset{\ask{Q'_j}{d}}{\ask{Q_j}{d} \in \gamma, j \in \set{\range{1}{n}}%, i \neq j
}$. We show that, indeed, $(\arule{l}{\alpha}) \in Q^M_i$ reappears in $(Q'_i+)^{M'}$ under the form $(\arule{l}{\alpha'})$.

First, whenever $\ask{Q_k}{c} \in \beta$, we know that $\ask{Q_k}{c} \notin M$ since $\beta \cap M = \emptyset$. From the construction of $M'$ we have that $\ask{N_k}{\neg\fresh{c}} \in M'$. Similarly, since $\gamma \subseteq M$ we know from the construction of $M'$ that $\ask{Q'_j}{d} \in M'$ whenever $\ask{Q_j}{d} \in \gamma$. Hence when determining the reduct~${(Q'_i+)}^{M'}$, the extended situated literals in $\beta'$ and $\gamma'$ will be deleted.

Finally, whenever $\alpha \cap M \neq \emptyset$ we know from the construction of $M'$ that $\ask{Q'_i}{b} \in M'$ whenever $\ask{Q_i}{b} \in \alpha$. Clearly, when determining the reduct, none of these extended situated literals will be deleted as they are $Q'_i$-local. Hence it is clear that the reduct of the communicating rule (\arule{\ask{Q'_i}{l}}{\alpha' \cup \beta' \cup \gamma'}) is the rule \arule{\ask{Q'_i}{l}}{\alpha'}. This completes the first part of the proof. As indicated, the second part of the proof is completely analogous. 
\end{proof}

\begin{reproposition}{partialanswerset}
Let $\struprog{P} = \set{\srange{Q}{1}{n}}$ and let $\struprog{P}' = \set{\srange{Q'}{1}{n}, \srange{N}{1}{n}}$ with $\struprog{P}$ a communicating normal program and $\struprog{P}'$ the communicating simple program that simulates \struprog{P} as defined in \themydef{simulateextended}. If $M$ is an answer set of \struprog{P}, then $M'$ is an answer set of $\struprog{P}'$ with $M'$ defined as in \themylem{analogousanswerset}.
\end{reproposition}
\begin{proof}
This proof is divided into two parts. In part 1 we only consider the component programs $Q'_i$ with $i \in \set{\range{1}{n}}$ and show that ${\left((Q'_i)^{M'}\right)}^{\star} = \proj{(M')}{Q'_i}$. In part 2 we do the same, but we only consider the component programs $N_i$ with $i \in \set{\range{1}{n}}$. As per \themydef{answerset-communicating} we have then shown that $M'$ is indeed an answer set of $\struprog{P}'$.

Consider a component program $Q'_i$ with $i \in \set{\range{1}{n}}$. By \themydef{simulateextended} we have that $Q'_i = (Q'_i+) \cup (Q'_i-)$ and thus
 \begin{equation}\label{prop-part1}
   {(Q'_i)}^{M'} = {(Q'_i+)}^{M'} \cup {(Q'_i-)}^{M'}.
 \end{equation}
For ${Q'_i-}$ we know by construction that it only contains rules that are of the form $(\arule{\ask{Q'_i}{\neg\fresh{b}}}{\ask{N_i}{\neg\fresh{b}}})$ and that the only rules of this form are in $Q'_i-$. Therefore, due to the definition of the reduct, we have
\begin{align}
  {(Q'_i-)}^{M'} &= \condset{\arule{\neg\fresh{b}}{}}{\ask{N_i}{\neg\fresh{b}} \in M'}\nonumber\\
  \intertext{and because of the construction of $M'$, see \Ref{analogousanswerset}, we obtain}
  {(Q'_i-)}^{M'} &= \condset{\arule{\neg\fresh{b}}{}}{b \notin \proj{M}{Q_i}}.\label{prop-part2}
\end{align}
Hence ${(Q'_i-)}^{M'}$ only contains facts about literals that, by construction of $Q'_i$, do not occur in $Q'_i+$. This means that from \Ref{prop-part1} and \Ref{prop-part2} we obtain
 \begin{equation}
   {\left({(Q'_i)}^{M'}\right)}^{\star} = {\left({(Q'_i+)}^{M'}\right)}^{\star} \cup \condset{\arule{\neg\fresh{b}}{}}{b \notin \proj{M}{Q_i}}\label{prop-part3}.
 \end{equation}
From \themylem{analogousanswerset} we know that $(Q'_i+)^{M'} = \condset{\arule{l}{\alpha'}}{\arule{l}{\alpha} \in Q^M_i}$ where $\alpha' = \condset{\ask{Q'_i}{b}}{\ask{Q_i}{b} \in \alpha}$. Because of the definition of an answer set of a communicating program we have
 \begin{equation}
   \proj{M}{Q_i} = {\left(Q_i^{M}\right)}^{\star} = {\left({(Q'_i+)}^{M'}\right)}^{\star}\label{prop-part4}.
 \end{equation}
Combining this with \eqref{prop-part3} we get
 \begin{align*}
      {\left({(Q'_i)}^{M'}\right)}^{\star} &= \proj{M}{Q_i} \cup \condset{\neg\fresh{b}}{b \notin \proj{M}{Q_i}}\\
      &= \proj{(M')}{Q'_i}&&\text{(definition of $M'$, see \Ref{analogousanswerset})}
 \end{align*}
This concludes the first part of the proof.

In the second part of the proof, we only consider the component programs $N'_i$ with $i \in \set{\range{1}{n}}$. By construction of $N_i$ we know that all the rules of the form $\arule{\neg\fresh{b}}{\ask{Q'_i}{\neg\fresh{b}}}$ and $\arule{\fresh{b}}{\ask{Q'_i}{b}}$ are in $N_i$ and that all the rules in $N_i$ are of this form. We have
\begin{align*}
  ({N_i})^{M'} &=
  \condset{\arule{\neg\fresh{b}}{}}{\ask{Q'_i}{\neg\fresh{b}} \in M'} \cup \condset{\arule{\fresh{b}}{}}{\ask{Q'_i}{b} \in M'}
  \intertext{which, due to the definition of $M'$ can be written as}
  &= \condset{\arule{\neg\fresh{b}}{}}{b \notin \proj{M}{Q_i}} \cup \condset{\arule{\fresh{b}}{}}{b \in \proj{M}{Q_i}}
\end{align*}
from which it follows that ${\left(\left(N_i\right)^{M'}\right)}^{\star} = \proj{(M')}{N_i}.$
\end{proof}

%\begin{relemma}{pkiffnk}
%Let $\struprog{P}' = \set{\srange{Q'}{1}{n}, \srange{N}{1}{n}}$ be a communicating simple program that simulates a communicating program $\struprog{P} = \set{\srange{Q}{1}{n}}$ as defined in \themydef{simulateextended}. If $M'$ is an answer set of $\struprog{P}'$, we must have that $\left(\ask{Q'_k}{\neg\fresh{ c}}\right) \in M'$ if and only if $\left(\ask{N_k}{\neg\fresh{ c}}\right) \in M'$ for any $k$ where $1 \leq k \leq n$.
%\end{relemma}
%\begin{proof}
%This follows straightforwardly from the fact that $(\arule{\neg\fresh{c}}{}) \in {(Q'_k)}^{M'}$ as soon as $(\ask{N_k}{\neg\fresh{c}}) \in M'$ and conversely $(\arule{\neg\fresh{c}}{}) \in {(N_k)}^{M'}$ as soon as $(\ask{Q_k}{\neg\fresh{c}}) \in M'$.
%\end{proof}

\begin{relemma}{reverseanalogousanswerset}
Let $\struprog{P} = \set{\srange{Q}{1}{n}}$ and let $\struprog{P}' = \set{\srange{Q'}{1}{n}, \srange{N}{1}{n}}$ with $\struprog{P}$ a communicating normal program and $\struprog{P'}$ the communicating simple program that simulates \struprog{P}. Assume that $M'$ is an answer set of $\struprog{P'}$ and that $\proj{(M')}{N_i}$ is total \wrt $\hbase{N_i}$ for all $i \in \set{\range{1}{n}}$. Let $M$ be defined as
  \begin{align}
    M = & \condset{\ask{Q_i}{b}}{\ask{Q'_i}{b} \in \left(\left(Q'_i+\right)^{M'}\right)^{\star}}
  \end{align}
  For each $i \in \set{\range{1}{n}}$, it holds that $(Q'_i+)^{M'} = \condset{\arule{l}{\alpha'}}{\arule{l}{\alpha} \in Q^M_i}$ with $\alpha' = \condset{\ask{Q'_{i}}{b}}{\ask{Q_{i}}{b} \in \alpha}$.
\end{relemma}
\begin{proof}
To prove this, we first show that any rule of the form $(\arule{l}{\alpha'}) \in {(Q'_i+)}^{M'}$ reappears in $Q^M_i$ under the form $\arule{l}{\alpha}$ for any $i \in \set{\range{1}{n}}$. We then show that the converse also holds, which is rather analogous to the proof  of the first part of \themylem{analogousanswerset}. Due to some technical subtleties in the second part of the proof, however, we present the proof in detail.

Suppose $(\arule{l}{\alpha'}) \in {(Q'_i+)}^{M'}$. By the definition of the reduct of a communicating simple program we know that there is some communicating simple rule of the form $(\arule{l}{\alpha' \cup \beta' \cup \gamma'}) \in Q'_i+$ such that $\beta' \subseteq M'$ is a set of situated literals of the form $\ask{N_k}{\neg\fresh{ c}}$ and $\gamma' \subseteq M'$ is a set of situated literals of the form $\ask{Q'_j}{d}$ with $i \neq j$ and $1 \leq j,k \leq n$.

From the definition of $Q'_i+$, we know that $(\arule{\ask{Q'_i}{l}}{\alpha' \cup \beta' \cup \gamma'})$ corresponds to a rule $(\arule{l}{\alpha \cup \naf \beta \cup \gamma}) \in Q_i$ where we have that $\alpha = \condset{\ask{Q_i}{b}}{\ask{Q'_i}{b} \in \alpha'}$, $\beta = \condset{\ask{Q_k}{c}}{\ask{N_k}{\neg\fresh{ c}} \in \beta'}$ and $\gamma = \condset{\ask{Q_j}{d}}{\ask{Q'_j}{d} \in \gamma'%, i \neq j
}$. We show that, indeed, $(\arule{l}{\alpha'}) \in {(Q'_i+)}^{M'}$ reappears in $Q^M_i$ under the form $(\arule{l}{\alpha})$.

First, since $\beta' \subseteq M'$, whenever $\ask{N_k}{\neg\fresh{c}} \in \beta'$ we know that $\ask{N_k}{\neg\fresh{c}} \in M'$. Since $M'$ is a model (indeed, it is an answer set) it is an interpretation (and thus consistent). Therefore, if $\ask{N_k}{\neg\fresh{c}} \in M'$ then surely $\ask{N_k}{\fresh{c}} \notin M'$. Now, if we were to have $\ask{Q'_k}{c} \in M'$, then applying the immediate consequence operator on the rule $\arule{\ask{N_k}{\fresh{c}}}{\ask{Q'_k}{c}}$ found in the component program $N_k$ would force us to have $\ask{N_k}{\fresh{c}} \in M'$ which results in a contradiction. Hence we find that $\ask{Q'_k}{c} \notin M'$. By \themydef{simulateextended} we know that $Q'_k = (Q'_k+) \cup (Q'_k-)$ and thus, by the definition of the reduct, we know that ${(Q'_k)}^{M'} = {(Q'_k+)}^{M'} \cup {(Q'_k-)}^{M'}$. Then we find that ${({(Q'_k)}^{M'})}^{\star} = {({(Q'_k+)}^{M'})}^{\star} \cup {({(Q'_k-)}^{M'})}^{\star}$ since all the rules in $Q'_k-$ have fresh literals in the head and literals from $N_k$ in the body and hence cannot interact with the rules from $Q'_k+$ which only depend on information derived from $Q'_k+$ and $N_k$ in their bodies. Recall from the definition of an answer set of a communicating program that $\forall Q'_k \in \struprog{P'} \cdot (\ask{Q'_k}{\proj{M'}{Q'_k}}) = {\left({(Q'_k)}^{M'}\right)}^{\star}$. Since we already found that $\ask{Q'_k}{c} \notin M'$ %and because $Q'_k-$ only contains rules with fresh literals in the head,
 we must have $\ask{Q'_k}{c} \notin \left(\left(Q'_i+\right)^{M'}\right)^{\star}$, or, because of the definition of $M$, that $\ask{Q_k}{c} \notin M$. Hence when determining the reduct ${(\arule{l}{\alpha \cup \naf \beta \cup \gamma})}^{M}$, the extended situated literals in $\naf \beta$ will be deleted.

In a similar way of reasoning, since $\gamma' \subseteq M'$ and because $\gamma = \condset{\ask{Q_j}{d}}{\ask{Q'_j}{d} \in \gamma'}$ we know from the construction of $M$ that $\gamma \subseteq M$. Hence when determining the reduct, the situated literals in $\gamma$ will be deleted. Finally, since $\alpha' \subseteq M'$ and because $\condset{\ask{Q_i}{b}}{\ask{Q'_i}{b} \in \alpha'} \subseteq M$ we know from the construction of $M$ that $\alpha \in M$. Clearly, when determining the reduct, none of the situated literals in $\alpha$ will be deleted as they are $Q_i$-local. Hence the reduct of the communicating rule (\arule{\ask{Q_i}{l}}{\alpha \cup \naf \beta \cup \gamma}) is the rule \arule{\ask{Q_i}{l}}{\alpha}. This completes the first part of the proof. 

% As indicated, the second part of the proof is completely analogous. 

%%%%%%%%%%%
%%%%%%%%%%%
%%%%%%%%%%%
%%%%%%%%%%%
%%%%%%%%%%%
%%%%%%%%%%%

We now come to the second part. This time we show that any rule of the form $(\arule{l}{\alpha}) \in Q^M_i$ reappears in $(Q'_i+)^{M'}$ under the form $(\arule{l}{\alpha'})$ for any $i \in \set{\range{1}{n}}$. 

Suppose $(\arule{l}{\alpha}) \in Q^M_i$ for some $i \in \set{\range{1}{n}}$. By the definition of the reduct we know that there is some rule of the form $(\arule{l}{\alpha \cup \naf \beta \cup \gamma}) \in Q_i$ such that $\beta \cap M = \emptyset$ and $\gamma \subseteq M$ is a set of situated literals of the form $\ask{Q_j}{d}$ with $i \neq j$, $1 \leq j \leq n$. From \themydef{simulateextended}, we know that the communicating normal rule $(\arule{\ask{Q_i}{l}}{\alpha \cup \naf \beta \cup \gamma})$ is transformed into the rule $(\arule{\ask{Q'_i}{l}}{\alpha' \cup \beta' \cup \gamma'})$ with $\alpha' = \condset{\ask{Q'_i}{b}}{\ask{Q_i}{b} \in \alpha}$, $\beta' = \condset{\ask{N_k}{\neg\fresh{ c}}}{\ask{Q_k}{c} \in \beta, k \in \set{\range{1}{n}}}$ and $\gamma' = \condset{\ask{Q'_j}{d}}{\ask{Q_j}{d} \in \gamma, j \in \set{\range{1}{n}}%, i \neq j
}$. We show that, indeed, $(\arule{l}{\alpha}) \in Q^M_i$ reappears in $(Q'_i+)^{M'}$ under the form $(\arule{l}{\alpha'})$ when $\proj{M'}{N_i}$ is total \wrt $\hbase{N_i}$ for all $i \in \set{\range{1}{n}}$.

First, since $\gamma \subseteq M$ we know from the construction of $M$ that $\ask{Q_j}{d} \in M$ whenever $\ask{Q'_j}{d} \in \gamma'$. Also, when $\ask{Q_k}{c} \in \beta$, we know that $\ask{Q_k}{c} \notin M$ since $\beta \cap M = \emptyset$. From the construction of $M$ we then know that $\ask{Q'_k}{c} \notin M'$ and since $M'$ is an answer set we readily obtain that $\ask{N_k}{\fresh{c}} \notin M'$ due to the construction of $N_k$. Together with the requirement that $\proj{M'}{N_k}$ is total \wrt $\hbase{N_k}$ we then must have that $\ask{N_k}{\neg\fresh{c}} \in M'$. Hence when determining the reduct~${(Q'_i+)}^{M'}$, the extended situated literals in $\beta'$ and $\gamma'$ will be deleted.

Finally, whenever $\alpha \cap M \neq \emptyset$ we know from the construction of $M'$ that $\ask{Q'_i}{b} \in M'$ whenever $\ask{Q_i}{b} \in \alpha$. Clearly, when determining the reduct, none of these extended situated literals will be deleted as they are $Q'_i$-local. Hence it is clear that the reduct of the communicating rule (\arule{\ask{Q'_i}{l}}{\alpha' \cup \beta' \cup \gamma'}) is the rule \arule{\ask{Q'_i}{l}}{\alpha'}. This completes the second part of the proof.

\end{proof}

\begin{reproposition}{converse}
Let $\struprog{P} = \set{\srange{Q}{1}{n}}$ and let $\struprog{P}' = \set{\srange{Q'}{1}{n}, \srange{N}{1}{n}}$ with $\struprog{P}$ a communicating normal program and $\struprog{P'}$ the communicating simple program that simulates \struprog{P}. Assume that $M'$ is an answer set of $\struprog{P'}$ and that $\proj{(M')}{N_i}$ is total \wrt $\hbase{N_i}$ for all $i \in \set{\range{1}{n}}$. Then the interpretation $M$ defined in \themylem{reverseanalogousanswerset} is an answer set of \struprog{P}. 
\end{reproposition}

\begin{proof}
  \themylem{reverseanalogousanswerset} tells us that $(Q'_i+)^{M'} = \condset{\arule{l}{\alpha'}}{\arule{l}{\alpha} \in Q^M_i}$ where we have $\alpha' = \condset{\ask{Q'_{i}}{b}}{\ask{Q_{i}}{b} \in \alpha}$. Hence we have ${\left((Q'_i+)^{M'}\right)}^{\star} = {\left(Q^M_i\right)}^{\star}$ since repeatedly applying the immediate consequence operator must conclude the same literals~$l$ due to the correspondence of the rules in the reducts and because of the way $\alpha'$ is defined. Since we defined $M$ as
 $$
     \condset{\ask{Q_i}{b}}{\ask{Q'_i}{b} \in \left(\left(Q'_i+\right)^{M'}\right)^{\star}}
 $$ 
   it follows immediately that $M$ is an answer set of \struprog{P} since
  \begin{equation}
    \forall i \in \set{\range{1}{n}} \cdot {\left(Q^M_i\right)}^{\star} = \proj{M}{Q_i}
  \end{equation}
  which completes the proof.
\end{proof}

\begin{reproposition}{one-focused-is-polynomial}
Let \struprog{P} be a communicating simple program. We then have:
\begin{list}{\labelitemi}{\leftmargin=2em}
  \item there always exists at least one $\vect{Q_1, ..., Q_n}$-focused answer set of $\struprog{P}$;
  \item we can compute this $\vect{Q_1, ..., Q_n}$-focused answer set in polynomial time. 
\end{list}
\end{reproposition}
\begin{proof}
We know from \themyprop{one-simple-is-polynomial} that we can always find a globally minimal answer of $\struprog{P}$ in polynomial time. Due to the way we defined the immediate fixpoint operator $T_{\struprog{P}}$ this operator only makes true the information that is absolutely necessary, \ie the minimal amount of information that can be derived (for each component program). It is then easy to see that no component program can derive any less information (we have no negation-as-failure) and thus that this globally minimal answer set is also locally minimal and thus a $\vect{Q_1, ..., Q_n}$-focused answer set of $\struprog{P}$. 
\end{proof}
%\begin{proof}
%We know from \themyprop{one-simple-is-polynomial} that computing one of the answer sets, say $M$, of a communicating simple program \struprog{P} can be done in polynomial time. In the proof of \themyprop{one-simple-is-polynomial}, we showed this by  introducing a generalised immediate consequence operator $T_{\struprog{P}}$ for the case of communicating simple programs. The least (informative) fixpoint $M$ of this generalised operator $T_{\struprog{P}}$ is an answer set of \struprog{P}. 
%
%In order to prove that we can compute the $\vect{\srange{Q}{1}{n}}$-focused answer sets of $\struprog{P}$ in polynomial time, we need to show that the answer set $M$ is minimal for all component programs $Q_i$, $1 \leq i \leq n$. This can readily be seen. The conclusions derived by the operator $T_{\struprog{P}}$ are the conclusions obtained by starting from the empty interpretation, \ie all the conclusions that follow readily from the facts in the communicating simple program. Furthermore, since the rules do not contain negation-as-failure, none of these conclusions can be retracted when more information becomes known. It is hence not possible to derive any less information. In other words, for all component programs $Q_i$ and all answer sets $M'$ of \struprog{P}, we must have that $\proj{M'}{Q_i} \not\subset \proj{M}{Q_i}$ and thus $M$ is a $\vect{\srange{Q}{1}{n}}$-focused answer sets of $\struprog{P}$.
%\end{proof}

\begin{reproposition}{qbf}
  Let $\phi$ and \struprog{P} be as in \themydef{simulation-QBF}.
  We have that a QBF $\phi$ of the form $\phi = \exists X_1 \forall X_2 ... \mathrm{\Theta} X_n \cdot p(X_1, X_2, \cdots X_n)$ is satisfiable if and only if $\ask{Q_0}{sat}$ is true in some $\vect{\srange{Q}{1}{n-1}}$-focused answer set of $\struprog{P}$.  Furthermore, we have that a QBF $\phi$ of the form $\phi = \forall X_1 \exists X_2 ... \mathrm{\Theta} X_n \cdot p(X_1, X_2, \cdots X_n)$ is satisfiable if and only if $\ask{Q_0}{sat}$ is true in all $\vect{\srange{Q}{1}{n-1}}$-focused answer sets of $\struprog{P}$. 
\end{reproposition}

\begin{proof}
We give a proof by induction. Assume we have a QBF $\phi_1$ of the form $\exists X_1 \cdot p(X_1)$ with $\struprog{P}_1 = \set{Q_0}$ the communicating normal program corresponding with $\phi_1$ according to~\themydef{simulation-QBF}. If the formula $p_1(X_1)$ of the QBF $\phi_1$ is satisfiable then we know that there is a $\vect{}$-focused answer set $M$ of $\struprog{P}_1$ such that $\ask{Q_0}{sat} \in M$. Otherwise, we know that $\ask{Q_0}{sat} \notin M$ for all $\vect{}$-answer sets $M$ of $\struprog{P}_1$. Hence the induction hypothesis is valid for $n=1$.

Assume the result holds for any QBF $\phi_{n-1}$ of the form $\exists X_1 \forall X_2 \ldots \mathrm{\Theta} X_{n-1} \cdot p_n(X_1, \srange{X}{2}{n-1})$. We show in the induction step that it holds for any QBF $\phi_{n}$ of the form $\exists X_1 \forall X_2 \ldots \overline{\mathrm{\Theta}} X_{n} \cdot p_{n-1}(X_1, \srange{X}{2}{n})$. Let $\struprog{P} = \set{\srange{Q}{0}{n-1}}$ and $\struprog{P}' = \set{\srange{Q'}{0}{n-2}}$ be the communicating normal programs that correspond with $\phi_{n}$ and $\phi_{n-1}$, respectively. Note that the component programs $\srange{Q}{2}{n-1}$ are defined in exactly the same way as the component programs $\srange{Q'}{1}{n-2}$, the only difference being the name of the component programs. What is of importance in the case of $\phi_n$ is therefore only the additional rules in $Q_0$ and the new component program $Q_1$. The additional rules in $Q_0$ merely generate the corresponding interpretations, where we now need to consider the possible interpretations of the variables from $X_n$ as well. The rules in the new component program $Q_1$ ensure that $\ask{Q_1}{x} \in M$ whenever $\ask{Q_0}{x} \in M$ and $\ask{Q_1}{\neg x} \in M$ whenever $\ask{Q_0}{\neg x} \in M$ for every $M$ an answer set of \struprog{P} and $x \in (\srangec{X}{1}{n-1}{\cup})$. Depending on $n$ being even or odd, we get two distinct cases:

\begin{list}{\labelitemi}{\leftmargin=2em}
  \item if $n$ is even, then we have $(\arule{sat}{\ask{Q_0}{sat}}) \in Q_1$ and we know that the QBF~$\phi_n$ has the form $\exists X_1 \forall X_2 \ldots \forall X_{n} \cdot p_n(X_1, \srange{X}{2}{n})$. Let us consider what happens when we determine the $\vect{Q_1}$-focused answer sets of \struprog{P}. Due to the construction of $Q_1$, we know that $\proj{M'}{Q_1} \subset \proj{M}{Q_1}$ can only hold for two answer sets $M'$ and $M$ of \struprog{P} if $M'$ and $M$ correspond to identical interpretations of the variables in $\srangec{X}{1}{n-1}{\cup}$. Furthermore, $\proj{M'}{Q_1} \subset \proj{M}{Q_1}$ is only possible if $\ask{Q_1}{sat} \in M$ while $\ask{Q_1}{sat} \notin M'$. 
  
  Now note that given an interpretation of the variables in $\srangec{X}{1}{n-1}{\cup}$, there is exactly one answer set for each choice of $X_n$. When we have $M'$ with $\ask{Q_1}{sat} \notin M'$ this implies that there is an interpretation such that, for some choice of $X_n$, this particular assignment of values of the QBF does not satisfy the QBF. Similarly, if we have $M$ with $\ask{Q_1}{sat} \in M$ then the QBF is satisfied for that particular choice of $X_n$. Determining $\vect{Q_1}$-focused answer sets of \struprog{P} will eliminate $M$ since $\proj{M'}{Q_1} \subset \proj{M}{Q_1}$. In other words, for identical interpretations of the variables in $\srangec{X}{1}{n-1}{\cup}$, the answer set $M'$ encodes a counterexample that shows that for these interpretations it does not hold that the QBF is satisfied for all choices of $X_n$. Focusing thus eliminates those answer sets that claim that the QBF is satisfiable for the variables in $\srangec{X}{1}{n-1}{\cup}$. When we cannot find such $\proj{M'}{Q_1} \subset \proj{M}{Q_1}$ this is either because none of the interpretations satisfy the QBF or all of the interpretations satisfy the QBF. In both cases, there is no need to eliminate any answer sets. We thus effectively mimic the requirement that the QBF $\phi_n$ should hold for $\forall X_n$.
  \\
  \item if $n$ is odd, then $(\arule{\neg sat}{\ask{Q_0}{\neg sat}}) \in Q_1$ and we know that the QBF $\phi_n$ has the form $\exists X_1 \forall X_2 \ldots \exists X_{n} \cdot p_n(X_1, \srange{X}{2}{n})$. As before, we know that $\proj{M'}{Q_1} \subset \proj{M}{Q_1}$ can only hold for two answer sets $M'$ and $M$ of \struprog{P} if $M'$ and $M$ correspond to identical interpretations of the variables in $\srangec{X}{1}{n-1}{\cup}$. However, this time $\proj{M'}{Q_1} \subset \proj{M}{Q_1}$ is only possible if $\ask{Q_1}{\neg sat} \in M$ while $\ask{Q_1}{\neg sat} \notin M'$. 
  
  If we have $M$ with $\ask{Q_1}{\neg sat} \in M$ then the QBF is not satisfied for that particular choice of $X_n$, whereas when $M'$ with $\ask{Q_1}{\neg sat} \notin M'$ this implies that there is an interpretation such that, for some choice of $X_n$, this particular assignment of the variables does satisfy the QBF. Determining $\vect{Q_1}$-focused answer sets of \struprog{P} will eliminate $M$ since $\proj{M'}{Q_1} \subset \proj{M}{Q_1}$. For identical interpretations of the variables in $\srangec{X}{1}{n-1}{\cup}$, the answer set $M'$ encodes a counterexample that shows that for these interpretations there is some choice of $X_n$ such that the QBF is satisfied. Focusing thus eliminates those answer sets that claim that the QBF is not satisfiable for the variables in $\srangec{X}{1}{n-1}{\cup}$. When we cannot find such $\proj{M'}{Q_1} \subset \proj{M}{Q_1}$ this is either because none of the interpretations satisfy the QBF or all of the interpretations satisfy the QBF. In both cases, there is no need to eliminate any answer sets. We effectively mimic the requirement that the QBF $\phi_n$ should hold for $\exists X_n$.
\end{list}

For a QBF of the form $\forall X_1 \exists X_2 \ldots \mathrm{\Theta} X_{n} \cdot p(X_1, \srange{X}{2}{n})$, with $\mathrm{\Theta} = \exists$ if $n$ is even and $\mathrm{\Theta} = \forall$ otherwise, the proof is analogous. In the base case, we know that a QBF $\phi_1$ of the form $\forall X_1 \cdot p(X_1)$ is satisfiable only when for every $\vect{}$-focused answer set $M$ of $\struprog{P}_1 = \set{Q_0}$ we find that $\ask{Q_0}{sat} \in M$. Otherwise, we know that there exists some $\vect{}$-focused answers sets $M$ of $\struprog{P}_1$ such that $\ask{Q_0}{sat} \notin M$. Hence the induction hypothesis is valid for $n=1$. The induction step is then entirely analogous to what we have proven before, with the only difference being that the cases for $n$ being even or odd are swapped. Finally, since the first quantifier is $\forall$, we need to verify that $\ask{Q_0}{sat}$ is true in every $\vect{\srange{Q}{1}{n-1}}$-focused answer set of $\struprog{P}$.
\end{proof}

\begin{reproposition}{focused-complete}
  Let \struprog{P} be a communicating normal program with $Q_i \in \struprog{P}$. The problem of deciding whether there exists a $\vect{\srange{Q}{1}{n}}$-focused answer set $M$ of $\struprog{P}$ such that  $\ask{Q_i}{l} \in M$ (brave reasoning) is in \spolsig{n+1}.
\end{reproposition}
\begin{proof}
We show the proof by induction on $n$. In the case where $n=1$, we need to guess a $\vect{Q_1}$-focused answer set $M$ of $\struprog{P}$ which can clearly be done in polynomial time. We now need to verify that this is indeed a $\vect{Q_1}$-focused answer set which is a problem in $\ccoNP$. Indeed, verifying that $M$ is not a $\vect{Q_1}$-focused answer set can be done using the following procedure in \cNP:
\begin{list}{\labelitemi}{\leftmargin=2em}
  \item guess an interpretation $M'$
  \item verify that $M'$ is an answer set of \struprog{P}
  \item verify that $\proj{M'}{Q_1} \subset \proj{M}{Q_1}$.
\end{list}
Hence, to find a \vect{Q_1}-focused answer set, we guess an interpretation, verify that it is an answer set in polynomial time, and we subsequently use an \cNP\ oracle to decide whether this answer set is \vect{Q_1}-focused, \ie the problem is in \spolsig{2}.
Assume that there exists an algorithm to compute the $\vect{\srange{Q}{1}{n-1}}$-focused answer sets of \struprog{P} that is in $\spolsig{n}$. In a similar fashion, we can guess a $\vect{\srange{Q}{1}{n}}$-focused answer set and verify there is no $\vect{\srange{Q}{1}{n}}$-focused answer set $M'$ of $\struprog{P}$ such that $\proj{M'}{Q_{n}} \subset \proj{M}{Q_{n}}$ using a \spolsig{n} oracle, \ie the algorithm is in $\spolsig{n+1}$.
\end{proof}

\begin{reproposition}{simple-complete}
 Let \struprog{P} be a communicating simple program with $Q_i \in \struprog{P}$. The problem of deciding whether there exists a $\vect{\srange{Q}{1}{n}}$-focused answer set $M$ of $\struprog{P}$ such that $\ask{Q_i}{l} \in M$ (brave reasoning) is in \spolsig{n+1}.
\end{reproposition}
\begin{proof}
We know from~\themyprop{partialanswerset} that one normal program can be simulated by a communicating simple program with two component programs. Since only the program $Q_0$ in the simulation in~\themydef{simulation-QBF} includes negation-as-failure, it suffices to add a single simple component program in order to simulate the negation-as-failure. Since the number of component programs is of no importance in~\themyprop{focused-complete}, the result readily follows.
\end{proof}

\begin{reproposition}{disjunctive-complete}
 Let \struprog{P} be a communicating disjunctive program with $Q_i \in \struprog{P}$. The problem of deciding whether $\ask{Q_i}{l} \in M$ with $M$ a $\vect{\srange{Q}{1}{n}}$-focused answer set of $\struprog{P}$ is in \spolsig{n+2}.
\end{reproposition}
\begin{proof}
This result can easily be verified by looking at the proof of \themyprop{focused-complete} and noticing that the only part of the algorithm that is affected by the use of communicating disjunctive programs \struprog{P} is the base step. In this base step, we use an oracle in $\cNP$ to check whether our guess $M$ is indeed an answer set of $\struprog{P}$. Since $M$ is an answer set of $\struprog{P}$ iff $\forall i \in \set{\range{1}{n}} \cdot (Q_i^M)^\star = \proj{M}{Q_i}$ and since $Q_i$ is a disjunctive component program we know that we will instead need an oracle in $\spolsig{2}$ to deal with communicative disjunctive programs. The remainder of the algorithm sketch remains unaffected.
\end{proof}

\label{lastpage}

\begin{thebibliography}{}

\bibitem[\protect\citeauthoryear{Baral}{Baral}{2003}]{baral:knowledge}
{\sc Baral, C.} 2003.
\newblock {\em Knowledge, Representation, Reasoning and Declarative Problem
  Solving}.
\newblock Cambridge University Press.

\bibitem[\protect\citeauthoryear{Bauters}{Bauters}{2011}]{bauters:modeling}
{\sc Bauters, K.} 2011.
\newblock Modeling coalition formation using multi-focused answer sets.
\newblock In {\em Proceedings of ESSLLI'11 Student Session}.

\bibitem[\protect\citeauthoryear{Bauters, Janssen, Schockaert, Vermeir, and {De
  Cock}}{Bauters et~al\mbox{.}}{2010}]{bauters:communicating}
{\sc Bauters, K.}, {\sc Janssen, J.}, {\sc Schockaert, S.}, {\sc Vermeir, D.},
  {\sc and} {\sc {De Cock}, M.} 2010.
\newblock Communicating answer set programs.
\newblock In {\em Technical Communications of ICLP 2010}. Vol.~7. 34--43.

\bibitem[\protect\citeauthoryear{Bauters, Schockaert, Vermeir, and {De
  Cock}}{Bauters et~al\mbox{.}}{2011}]{bauters:communicatingpolynomial}
{\sc Bauters, K.}, {\sc Schockaert, S.}, {\sc Vermeir, D.}, {\sc and} {\sc {De
  Cock}, M.} 2011.
\newblock Communicating {ASP} and the polynomial hierarchy.
\newblock To appear in Proceedings of LPNMR-11.

\bibitem[\protect\citeauthoryear{Brain and {De Vos}}{Brain and {De
  Vos}}{2003}]{brain:oclp}
{\sc Brain, M.} {\sc and} {\sc {De Vos}, M.} 2003.
\newblock Implementing {OCLP} as a front-end for answer set solvers: From
  theory to practice.
\newblock In {\em Proceedings of ASP'05}.

\bibitem[\protect\citeauthoryear{Brewka and Eiter}{Brewka and
  Eiter}{2007}]{brewka:equilibria}
{\sc Brewka, G.} {\sc and} {\sc Eiter, T.} 2007.
\newblock Equilibria in heterogeneous nonmonotonic multi-context systems.
\newblock In {\em Proceedings of AAAI'07}. 385--390.

\bibitem[\protect\citeauthoryear{Brewka, Eiter, Fink, and Weinzierl}{Brewka
  et~al\mbox{.}}{2011}]{brewka:managed}
{\sc Brewka, G.}, {\sc Eiter, T.}, {\sc Fink, M.}, {\sc and} {\sc Weinzierl,
  A.} 2011.
\newblock Managed multi-context systems.
\newblock In {\em Proceedings of IJCAI'11}. 786--791.

\bibitem[\protect\citeauthoryear{Brewka, Roelofsen, and Serafini}{Brewka
  et~al\mbox{.}}{2007}]{brewka:contextual}
{\sc Brewka, G.}, {\sc Roelofsen, F.}, {\sc and} {\sc Serafini, L.} 2007.
\newblock Contextual default reasoning.
\newblock In {\em Proceedings of IJCAI'07}. 268--273.

\bibitem[\protect\citeauthoryear{Buccafurri, Caminiti, and Laurendi}{Buccafurri
  et~al\mbox{.}}{2008}]{buccafurri:logic}
{\sc Buccafurri, F.}, {\sc Caminiti, G.}, {\sc and} {\sc Laurendi, R.} 2008.
\newblock A logic language with stable model semantics for social reasoning.
\newblock In {\em Proceedings of ICLP'08}. 718--723.

\bibitem[\protect\citeauthoryear{Bylander}{Bylander}{1994}]{bylander:computational}
{\sc Bylander, T.} 1994.
\newblock The computational complexity of propositional {STRIPS} planning.
\newblock {\em Artificial Intelligence\/}~{\em 69}, 165--204.

\bibitem[\protect\citeauthoryear{Dao-Tran, Eiter, Fink, and
  Krennwallner}{Dao-Tran et~al\mbox{.}}{2009}]{dao-tran:modular}
{\sc Dao-Tran, M.}, {\sc Eiter, T.}, {\sc Fink, M.}, {\sc and} {\sc
  Krennwallner, T.} 2009.
\newblock Modular nonmonotonic logic programming revisited.
\newblock In {\em Proceedings of ICLP'09}. 145--159.

\bibitem[\protect\citeauthoryear{Dao-Tran, Eiter, Fink, and
  Krennwallner}{Dao-Tran et~al\mbox{.}}{2010}]{daotran:distributed}
{\sc Dao-Tran, M.}, {\sc Eiter, T.}, {\sc Fink, M.}, {\sc and} {\sc
  Krennwallner, T.} 2010.
\newblock Distributed nonmonotonic multi-context systems.
\newblock In {\em Proceedings of KR'10}.

\bibitem[\protect\citeauthoryear{{De Vos}, Crick, Padget, Brain, Cliffe, and
  Needham}{{De Vos} et~al\mbox{.}}{2005}]{devos:laima}
{\sc {De Vos}, M.}, {\sc Crick, T.}, {\sc Padget, J.}, {\sc Brain, M.}, {\sc
  Cliffe, O.}, {\sc and} {\sc Needham, J.} 2005.
\newblock {LAIMA}: A multi-agent platform using ordered choice logic
  programming.
\newblock In {\em Proceedings of DALT'05}. 72--88.

\bibitem[\protect\citeauthoryear{Dell'Acqua, Sadri, and Toni}{Dell'Acqua
  et~al\mbox{.}}{1999}]{acqua:communicating}
{\sc Dell'Acqua, P.}, {\sc Sadri, F.}, {\sc and} {\sc Toni, F.} 1999.
\newblock Communicating agents.
\newblock In {\em Proceedings of MAS'99}.

\bibitem[\protect\citeauthoryear{Drescher, Eiter, Fink, Krennwallner, and
  Walsh}{Drescher et~al\mbox{.}}{2011}]{drescher:symmetry}
{\sc Drescher, C.}, {\sc Eiter, T.}, {\sc Fink, M.}, {\sc Krennwallner, T.},
  {\sc and} {\sc Walsh, T.} 2011.
\newblock Symmetry breaking for distributed multi-context systems.
\newblock In {\em Proceedings of ICLP'10}. Lecture Notes in Computer Science,
  vol. 6645. 26--39.

\bibitem[\protect\citeauthoryear{Eiter, Faber, Leone, and Pfeifer}{Eiter
  et~al\mbox{.}}{1999}]{eiter:diagnosis}
{\sc Eiter, T.}, {\sc Faber, W.}, {\sc Leone, N.}, {\sc and} {\sc Pfeifer, G.}
  1999.
\newblock The diagnosis frontend of the dlv system.
\newblock {\em AI Communications\/}~{\em 12}, 99--111.

\bibitem[\protect\citeauthoryear{Eiter and Gottlob}{Eiter and
  Gottlob}{1995}]{eiter:complexity}
{\sc Eiter, T.} {\sc and} {\sc Gottlob, G.} 1995.
\newblock The complexity of logic-based abduction.
\newblock {\em ACM\/}~{\em 42}, 3--42.

\bibitem[\protect\citeauthoryear{Eiter, Ianni, Lukasiewicz, Schindlauer, and
  Tompits}{Eiter et~al\mbox{.}}{2008}]{eiter:combining}
{\sc Eiter, T.}, {\sc Ianni, G.}, {\sc Lukasiewicz, T.}, {\sc Schindlauer, R.},
  {\sc and} {\sc Tompits, H.} 2008.
\newblock Combining answer set programming with description logics for the
  semantic web.
\newblock {\em Artifial Intelligence\/}~{\em 172,\/}~12--13, 1495--1539.

\bibitem[\protect\citeauthoryear{Eiter, Ianni, Schindlauer, and Tompits}{Eiter
  et~al\mbox{.}}{2005}]{eiter:uniform}
{\sc Eiter, T.}, {\sc Ianni, G.}, {\sc Schindlauer, R.}, {\sc and} {\sc
  Tompits, H.} 2005.
\newblock A uniform integration of higher-order reasoning and external
  evaluations in answer-set programming.
\newblock In {\em Proceedings of IJCAI'05}. 90--96.

\bibitem[\protect\citeauthoryear{Eiter, Ianni, Schindlauer, and Tompits}{Eiter
  et~al\mbox{.}}{2006}]{eiter:dlvhex}
{\sc Eiter, T.}, {\sc Ianni, G.}, {\sc Schindlauer, R.}, {\sc and} {\sc
  Tompits, H.} 2006.
\newblock dlvhex: A tool for semantic-web reasoning under the answer-set
  semantics.
\newblock In {\em Proceedings of International Workshop on Proceedings of
  ALPSWS'06}. 33--39.

\bibitem[\protect\citeauthoryear{Gebser, Guziolowski, Ivanchev, Schaub, Siegel,
  Thiele, and Veber}{Gebser et~al\mbox{.}}{2010}]{gebser:repair}
{\sc Gebser, M.}, {\sc Guziolowski, C.}, {\sc Ivanchev, M.}, {\sc Schaub, T.},
  {\sc Siegel, A.}, {\sc Thiele, S.}, {\sc and} {\sc Veber, P.} 2010.
\newblock Repair and prediction (under inconsistency) in large biological
  networks with answer set programming.
\newblock In {\em Proceedings of KR'10}.

\bibitem[\protect\citeauthoryear{Gelder, Ross, and Schlipf}{Gelder
  et~al\mbox{.}}{1991}]{gelder:well}
{\sc Gelder, A.~V.}, {\sc Ross, K.~A.}, {\sc and} {\sc Schlipf, J.~S.} 1991.
\newblock The well-founded semantics for general logic programs.
\newblock {\em Journal of the ACM\/}~{\em 38,\/}~3, 620--650.

\bibitem[\protect\citeauthoryear{Gelfond and Lifschitz}{Gelfond and
  Lifschitz}{1991}]{gelfond:classical}
{\sc Gelfond, M.} {\sc and} {\sc Lifschitz, V.} 1991.
\newblock Classical negation in logic programs and disjunctive databases.
\newblock {\em New Generation Computing\/}~{\em 9}, 365--385.

\bibitem[\protect\citeauthoryear{Gelfond and Lifzchitz}{Gelfond and
  Lifzchitz}{1988}]{gelfond:stablemodel}
{\sc Gelfond, M.} {\sc and} {\sc Lifzchitz, V.} 1988.
\newblock The stable model semantics for logic programming.
\newblock In {\em Proceedings of ICLP'88}. 1081--1086.

\bibitem[\protect\citeauthoryear{Giunchiglia and Serafini}{Giunchiglia and
  Serafini}{1994}]{giunchiglia:multilanguage}
{\sc Giunchiglia, F.} {\sc and} {\sc Serafini, L.} 1994.
\newblock Multilanguage hierarchical logics or: How we can do without modal
  logics.
\newblock {\em Artifial Intelligence\/}~{\em 65,\/}~1, 29--70.

\bibitem[\protect\citeauthoryear{Jeroslow}{Jeroslow}{1985}]{jeroslow:polynomial}
{\sc Jeroslow, R.} 1985.
\newblock The polynomial hierarchy and a simple model for competitive analysis.
\newblock {\em Mathematical Programming\/}~{\em 32}, 146--164.

\bibitem[\protect\citeauthoryear{Lifschitz}{Lifschitz}{2002}]{lifschitz:plan}
{\sc Lifschitz, V.} 2002.
\newblock Answer set programming and plan generation.
\newblock {\em Artificial Intelligence\/}~{\em 138}, 39--54.

\bibitem[\protect\citeauthoryear{Lifschitz, Tang, and Turner}{Lifschitz
  et~al\mbox{.}}{1999}]{lifschitz:nested}
{\sc Lifschitz, V.}, {\sc Tang, L.~R.}, {\sc and} {\sc Turner, H.} 1999.
\newblock Nested expressions in logic programs.
\newblock {\em Ann. Math. Artif. Intell.\/}~{\em 25,\/}~3-4, 369--389.

\bibitem[\protect\citeauthoryear{Luo, Shi, Wang, and Huang}{Luo
  et~al\mbox{.}}{2005}]{luo:cooperation}
{\sc Luo, J.}, {\sc Shi, Z.}, {\sc Wang, M.}, {\sc and} {\sc Huang, H.} 2005.
\newblock Multi-agent cooperation: A description logic view.
\newblock In {\em Proceedings of PRIMA'05}. 365--379.

\bibitem[\protect\citeauthoryear{Niemel\"{a} and Simons}{Niemel\"{a} and
  Simons}{2000}]{niemela:extending}
{\sc Niemel\"{a}, I.} {\sc and} {\sc Simons, P.} 2000.
\newblock Extending the smodels system with cardinality and weight constraints.
\newblock In {\em Logic-Based Artificial Intelligence}. Kluwer Academic
  Publishers, 491--521.

\bibitem[\protect\citeauthoryear{Papadimitriou}{Papadimitriou}{1994}]{papadimitriou:computational}
{\sc Papadimitriou, C.} 1994.
\newblock {\em Computational complexity}.
\newblock Addison-Wesley.

\bibitem[\protect\citeauthoryear{Roelofsen and Serafini}{Roelofsen and
  Serafini}{2005}]{roelofsen:minimal}
{\sc Roelofsen, F.} {\sc and} {\sc Serafini, L.} 2005.
\newblock Minimal and absent information in contexts.
\newblock In {\em Proceedings of IJCAI'05}. 558--563.

\bibitem[\protect\citeauthoryear{Tarski}{Tarski}{1955}]{tarski:lattice}
{\sc Tarski, A.} 1955.
\newblock A lattice-theoretical fixpoint theorem and its applications.
\newblock {\em Pacific Journal of Mathematics\/}~{\em 5,\/}~2, 285--309.

\bibitem[\protect\citeauthoryear{{Van Nieuwenborgh}, {De Vos}, Heymans, and
  Vermeir}{{Van Nieuwenborgh} et~al\mbox{.}}{2007}]{nieuwenborgh:hierarchical}
{\sc {Van Nieuwenborgh}, D.}, {\sc {De Vos}, M.}, {\sc Heymans, S.}, {\sc and}
  {\sc Vermeir, D.} 2007.
\newblock Hierarchical decision making in multi-agent systems using answer set
  programming.
\newblock In {\em Proceedings of CLIMA'07}.

\end{thebibliography}
\end{document}